\documentclass[review,12pt]{elsarticle}

\usepackage{amssymb}
\usepackage{amsthm}
\usepackage{physics}
\usepackage{amsmath}
\usepackage{mathrsfs}
\usepackage{graphicx}
\usepackage{epstopdf}
\usepackage{float}
\usepackage{color}
\usepackage{caption}
\usepackage{subcaption}
\usepackage{bm}
\usepackage{bbm}
\usepackage[euler]{textgreek}
\usepackage{hyperref}
\usepackage[capitalize]{cleveref}
\usepackage{soul}
\biboptions{sort&compress}
 \usepackage{lineno}
\usepackage{fancyhdr,lipsum}
\newcommand*\patchAmsMathEnvironmentForLineno[1]{%
	\expandafter\let\csname old#1\expandafter\endcsname\csname #1\endcsname
	\expandafter\let\csname oldend#1\expandafter\endcsname\csname end#1\endcsname
	\renewenvironment{#1}%
	{\linenomath\csname old#1\endcsname}%
	{\csname oldend#1\endcsname\endlinenomath}}%
\newcommand*\patchBothAmsMathEnvironmentsForLineno[1]{%
	\patchAmsMathEnvironmentForLineno{#1}%
	\patchAmsMathEnvironmentForLineno{#1*}}%
\AtBeginDocument{%
	\patchBothAmsMathEnvironmentsForLineno{equation}%
	\patchBothAmsMathEnvironmentsForLineno{align}%
	\patchBothAmsMathEnvironmentsForLineno{flalign}%
	\patchBothAmsMathEnvironmentsForLineno{alignat}%
	\patchBothAmsMathEnvironmentsForLineno{gather}%
	\patchBothAmsMathEnvironmentsForLineno{multline}%
}
 \crefname{appendix}{}{}

\journal{ }

\makeatletter
\def\@author#1{\g@addto@macro\elsauthors{\normalsize%
    \def\baselinestretch{1}%
    \upshape\authorsep#1\unskip\textsuperscript{%
      \ifx\@fnmark\@empty\else\unskip\sep\@fnmark\let\sep=,\fi
      \ifx\@corref\@empty\else\unskip\sep\@corref\let\sep=,\fi
      }%
    \def\authorsep{\unskip,\space}%
    \global\let\@fnmark\@empty
    \global\let\@corref\@empty  
    \global\let\sep\@empty}%
    \@eadauthor={#1}
}
\makeatother

\begin{document}

\begin{frontmatter}



\title{A coupled phase field formulation for modelling fatigue cracking in lithium-ion battery electrode particles}


\author{Weilong Ai\fnref{SEU,ICdyson,Faraday}}

\author{Billy Wu\fnref{ICdyson,Faraday}}

\author{Emilio Mart\'{\i}nez-Pa\~neda\corref{cor1}\fnref{IC}}
\ead{e.martinez-paneda@imperial.ac.uk}
\address[SEU]{School of Civil Engineering, Southeast University, Nanjing, China}
\address[ICdyson]{Dyson School of Design Engineering, Imperial College London, London SW7 2AZ, UK}

\address[Faraday]{The Faraday Institution, Quad One, Becquerel Avenue, Harwell Campus, Didcot, OX11 0RA, UK}

\address[IC]{Department of Civil and Environmental Engineering, Imperial College London, London SW7 2AZ, UK}

\cortext[cor1]{Corresponding author.}

\begin{abstract}
Electrode particle cracking is one of the main phenomena driving battery capacity degradation. Recent phase field fracture studies have investigated particle cracking behaviour. However, only the beginning of life has been considered and effects such as damage accumulation have been neglected. Here, a multi-physics phase field fatigue model has been developed to study crack propagation in battery electrode particles undergoing hundreds of cycles. In addition, we couple our electrochemo-mechanical formulation with X-ray CT imaging to simulate fatigue cracking of realistic particle microstructures. Using this modelling framework, non-linear crack propagation behaviour is predicted, leading to the observation of an exponential increase in cracked area with cycle number. Three stages of crack growth (slow, accelerating and unstable) are observed, with phenomena such as crack initialisation at concave regions and crack coalescence having a significant contribution to the resulting fatigue crack growth rates. The critical values of C-rate, particle size and initial crack length are determined, and found to be lower than those reported in the literature using static fracture models. Therefore, this work demonstrates the importance of considering fatigue damage in battery degradation models and provides insights on the control of fatigue crack propagation to alleviate battery capacity degradation.
\end{abstract}


\end{frontmatter}


\section{Introduction}
\label{Introduction}
Lithium-ion batteries are the main energy source for electric vehicles and consumer electronics. However, they suffer from capacity fade during their lifetime. One of the main degradation mechanisms is the fracture of electrode particles \cite{Palacin2016}, which is caused by the stresses associated with the inhomogeneous swelling and shrinkage of electrode materials that occurs when lithium-ions are inserted and extracted \cite{Lin2017}. The resulting cracks in the electrode particles lead to two negative effects on the battery performance: loss of electronic contact between particles, which decreases the amount of active material in a cell \cite{Zheng2012,Muller2018}, and additional parasitic side reactions that occur on fresh crack surfaces, e.g. the formation and growth of the solid electrolyte interphase (SEI), which leads to lithium inventory loss \cite{birkl2017}. Batteries often experience an accelerated degradation phase, where effects such as particle cracking become increasingly important as crack propagation rates increase with cycle number \cite{Deshpande2012,Edge2021}. Therefore, it is of great importance to account for particle cracking behaviour in lithium-ion battery models.

The modelling of mechanical effects in lithium-ion batteries needs consideration across multiple length scales, from particle level interactions at the micro-scale up to cell level effects at the macro-scale \cite{mukhopadhyay2014,zhang2017,Zhao2019a}. Early mechanical models for lithium-ion batteries include the work of Christensen and Newman \cite{Christensen2006,Christensen2006a} and Zhang et al. \cite{zhang2007} in the mid 2000s. These models were built to describe the volume expansion and stress generation during lithium (de)intercalation and diffusion in the electrode particles, where it was demonstrated that the stresses increase with particle size and current density \cite{Christensen2006,Christensen2006a,zhang2007}. In addition, the role of the hydrostatic stress in changing the chemical potential and accelerating lithium diffusion was quantified by Li et al. \cite{li2017}. The influence on the particle stress levels of electrode material properties such as the diffusion coefficient, the elastic modulus and the lithium partial molar volume has also been investigated \cite{Purkayastha2012}. By applying these electro-chemo-mechanical models at the cell level, the thickness change of a pouch cell during lithium (de)intercalation was found to be more than 10 times larger than that resulting from thermal expansion effects \cite{rieger2016c,Ai2020}. These cell-level coupled models have also been used to explain the experimentally observed localised particle fragmentation that takes place near the electrode-separator interface \cite{christensen2010}, which was attributed to the high reaction currents and large stresses close to the separator \cite{Ai2020}. 

Particle cracking phenomena have been frequently observed during battery cycling experiments \cite{ebner2013}, but high-fidelity modelling remains challenging because of the multiphysics nature of the problem. Mechanical stresses are a result of the chemical strains associated with lithium transport. During the lithium insertion process (lithiation), the centre of the electrode particle undergoes tensile stresses, while the outer regions are subjected to compression. The opposite occurs during the lithium extraction process (delithiation), as the outer surface layers of the particle are compressed during intercalation, while its interior is stretched. The magnitude of these stresses is strongly influenced by the particle geometry, where it has been shown that cracks have a higher propensity to initiate at the sharp corners of concave regions \cite{miehe2016}. Studies aiming at investigating these complex chemo-mechanical effects include Klinsmann et al. \cite{Klinsmann2016,Klinsmann2016a}, who used a phase field fracture model to capture crack growth in idealised particles. More recently, a similar model has been developed for describing complex crack paths in electrode particles, showing an approximate power law relationship between the critical flaw size for the onset of crack growth and the charging rate \cite{Mesgarnejad2019b}. Alternatively, Xu et al. \cite{xu2018} explored the use of cohesive zone models, where the crack path was defined \textit{a priori}. However, these models have only been used to study electrode particles under a single charge or discharge step while electrode particles are exposed to cyclic loading and experience fatigue damage. One attempt to account for fatigue crack growth is through the use of the Paris' law \cite{Paris1963} with 1D battery models \cite{Deshpande2012,Purewal2014}. However, this implies assuming that multiple, identical and equally-spaced micro cracks exist at the surfaces of idealised electrode particles, whose crack growth behaviour is at all times governed by Paris' law. Ekstr\"om and Lindbergh \cite{ekstrom2015} also proposed an empirical model to approximate the new surfaces resulting from crack propagation and the additional side reactions. The W\"ohler curve was used by Laresgoiti \textit{et al.} \cite{Laresgoiti2015} to estimate the capacity loss, considering that particle fracture leads to loss of active materials. Although those simplified models can be calibrated to capture battery capacity degradation up to 1 C currents, they fail in high C-rates, where the microstructure of electrode particles plays a more dominant role on battery performance. There is a need for a theoretical and numerical modelling framework capable of predicting fatigue crack growth in realistic particle geometries, under non-idealised stress states, and across all regimes of fatigue behaviour.    

In this work, we present the first multiphysics phase field model for capturing fatigue crack propagation in electrode particles. From this model, deeper understanding of the effects of C-rate, particle size and crack geometry on the fatigue cracks and the resulting battery degradation can be inferred. The theoretical foundations of the model are presented in \cref{Sec:NumModel}. The numerical implementation of this model, using the finite element method, is described in \cref{sec:FEimplementation}. Three boundary value problems are addressed in \cref{Sec:FEMresults} to study the fatigue crack behaviour of electrode particles under different working conditions. \cref{Sec:ConcludingRemarks} includes the main conclusions and findings of this work. The aim of this work is thus to provide both insights on fatigue cracking behaviour at the electrode particle level and perspectives towards linking these results to battery cell level performance.

\section{Phase field modelling of fatigue cracking in electrode particles}
\label{Sec:NumModel}

We proceed to describe our theory, which couples lithium diffusion (and the associated chemical strains, see Section \ref{Sec:LiDiffusion}), a phase field description of fracture (Section \ref{Sec:PhasefieldFracture}) and a variational fatigue damage model (Section \ref{Sec:FatigueDamage}).

\subsection{Lithium diffusion and chemical strains}
\label{Sec:LiDiffusion}

The coupled electrochemo-mechanical model for lithium diffusion in electrode particles was first proposed by Christensen and Newman \cite{Christensen2006,Christensen2006a} and Zhang et al. \cite{zhang2007}. This model was then applied to battery cells \cite{Ai2020}, to estimate the thickness change and the evolution of stress at different C-rates. Here only the key equations are included; more details can be found in Zhang et al. \cite{zhang2007}. 

The governing equation of lithium diffusion with no source terms is based on the conservation of lithium-ions via
\begin{subequations}\label{eq:diffusion_strong}
	\begin{gather}
	\derivative{c}{t}+\div{\textbf{J}}=0,\\
	\textbf{J}=-D\grad{c}+\frac{cD\Omega_\text{c}}{RT}\grad{\sigma_h} \quad \text{and} \quad \textbf{J}\cdot\textbf{n}=\bar{J}\quad \text{at } \partial \Omega,	
	\end{gather}
\end{subequations} 
where $c$ is the lithium concentration, $t$ is the time, $D$ is the diffusion coefficient, $\textbf{J}$ is the lithium flux vector, $\bar{J}$ is the magnitude of the external flux loading, $\Omega_\text{c}$ is the lithium partial molar volume, $R$ is the universal gas constant, $T$ is the temperature and $\sigma_h$ is the hydrostatic stress, i.e. $\sigma_h=\tr(\bm{\sigma})/3$. The chemical strain is 
determined by the local lithium concentration via
\begin{equation}
\bm{\varepsilon_\text{Li}} = \frac{\Omega_\text{c}(c-c_\text{0})}{3}\textbf{I},
\end{equation} 
where $c_0$, equal to zero in this work, is the reference lithium concentration in the stress-free state, and $\textbf{I}$ is the identity tensor. The constitutive model relating the Cauchy stress tensor $\bm{\sigma}$ to the strain tensor $\bm{\varepsilon}$ takes into consideration the chemical strain as follows:
\begin{equation}\label{eq:stress}
\bm{\sigma}= \lambda \tr(\bm{\varepsilon}-\bm{\varepsilon_\text{Li}})\textbf{I}+2\mu(\bm{\varepsilon}-\bm{\varepsilon_\text{Li}}),
\end{equation} 
where $\lambda$ and $\mu$ are the Lamé constants.    

\subsection{A phase field description of fracture}
\label{Sec:PhasefieldFracture}

Variational phase field fracture formulations have emerged as compelling methods for modelling crack nucleation and growth \cite{Bourdin2000,Tanne2018}. By using a scalar phase field $d$ to describe the crack-solid interface, phenomena such as crack deflection, the coalescence of multiple cracks and crack branching become easy to handle in arbitrary geometries and dimensions. Phase field methodologies have been successfully used to model fracture across a wide range of materials and applications, including composites \cite{Quintanas-Corominas2020,Tan2021}, functionally graded materials \cite{Hirshikesh2019,Kumar2021}, shape memory alloys \cite{Simoes2021,FFEMS2022}, rocks \cite{Wilson2016,Schuler2020}, and piezoelectric materials \cite{Abdollahi2012}. Phase field approaches have also been extended to coupled multi-physics problems of chemo-mechanical nature, such as hydrogen assisted fracture \cite{Martinez-Paneda2018,Anand2019,Wu2020a,Kristensen2020}, corrosion \cite{Mai2016,Cui2021}, cracking of nuclear fuel pellets \cite{Li2021} and particle fracture in Li-Ion batteries \cite{Klinsmann2016,Mesgarnejad2019b}. The phase field model for fracture has been extensively described elsewhere \cite{Bourdin2008,Kristensen2021} and thus only the main equations are introduced below.

In the phase field fracture model, a crack is treated as an interface between the intact and fully cracked areas, using $d=0$ and $d=1$, respectively. A smooth transition between the two states is assumed \cite{Miehe2010}. The evolution of the phase field (i.e., crack growth) follows Griffith theory, where the strain energy is released by creating new crack surfaces. The released strain energy equals the product of the cracked area and the critical energy release rate $G_\text{c}$, the material toughness. The governing equations are based on the equilibrium of stresses (with no body force) and a phase field evolution law based on Griffith's energy balance, using the so-called hybrid approach \cite{Ambati2014}:
\begin{subequations}\label{eq:stress_strong}
	\begin{gather}
	\div{[g(d)\bm{\sigma}]} = \textbf{0},\\
	\frac{G_\text{c}}{l}(d-l^2\laplacian{d})=2(1-d)\mathcal{H},\\
	\bm{\sigma}\cdot \textbf{n}=\bar{\textbf{t}} \quad \text{and} \quad \textbf{u}=\bar{\textbf{u}} \quad \text{at } \partial \Omega,\\
\grad{d}\cdot \textbf{n}=0 \quad \text{and} \quad d=0 \quad \text{at } \partial \Omega,	
	\end{gather}
\end{subequations}    
where $l$ is the characteristic phase field length, $\partial \Omega$ is the boundary with the outward normal $\textbf{n}$, $\bar{\textbf{t}}$ is the external force vector, $\textbf{u}$ is the displacement vector with constraints $\bar{\textbf{u}}$ and $g(d)$ is the phase field degradation function as
\begin{equation}\label{eq:degradation}
	g(d)=(1-d)^2 +k,
\end{equation}
with $k=10^{-5}$ to prevent ill-conditioning when $d=1$. $\mathcal{H}$ is the local history field of the maximum tensile elastic strain energy $\psi^+_0(t)$, $\mathcal{H}=\max \psi^+_0$. A volumetric-deviatoric split \cite{Amor2009} is used to prevent damage under compression, such that the the tensile and compressive strain energies are respectively defined as
\begin{subequations}
	\begin{gather}
	\psi^+_0=0.5K\big\langle\tr(\bm{\varepsilon}-\bm{\varepsilon_\text{Li}})\big\rangle_+^2 + \mu(\bm{\varepsilon}^\text{dev}:\bm{\varepsilon}^\text{dev}), \\
	 \psi^-_0  =0.5K\big\langle\tr(\bm{\varepsilon}-\bm{\varepsilon_\text{Li}})\big\rangle^2_-,
	\end{gather}
\end{subequations}
where $K$ is the bulk modulus, $\bm{\varepsilon}^\text{dev}$ is the deviatoric elastic strain, i.e. $\bm{\varepsilon}^\text{dev}=(\bm{\varepsilon}-\bm{\varepsilon_\text{Li}})-\tr(\bm{\varepsilon}-\bm{\varepsilon_\text{Li}})\textbf{I}/3$ and the operator $\langle \cdot \rangle_\pm$ is defined as $\langle x\rangle_\pm = (x\pm\abs{x})/2$. 

\subsection{Fatigue damage model}
\label{Sec:FatigueDamage}

Phase field fracture models have been very recently extended to model fatigue crack growth \cite{Lo2019,Carrara2020,Schreiber2020}. The accumulation of fatigue damage is introduced into the phase field fracture model by either adding a viscosity parameter to the energy balance equation \cite{Lo2019}, degrading the toughness using a fatigue damage function \cite{Carrara2020} or increasing the crack driving force \cite{Schreiber2020}. 

Here, we base our approach on the fatigue degradation function approach by  Carrara et al. \cite{Carrara2020}, which naturally recovers the Paris' law \cite{Paris1963} in the appropriate regime and under the appropriate conditions. Thus, the toughness $G_\text{c}$ is degraded in the presence of fatigue damage using a cumulative history variable $\bar{\alpha}$ and a degradation function $f(\bar{\alpha})$, i.e. the fatigue toughness is $G_\text{d}=f(\bar{\alpha})G_\text{c}$. Accordingly, the governing equation for the phase field is modified to
\begin{equation}\label{eq:fatigue_PF}
		\frac{G_\text{d}}{l}(d-l^2\laplacian{d})-l\grad{d}\cdot\grad{G_\text{d}}=2(1-d)\mathcal{H}.
\end{equation}
The variable $\bar{\alpha}$ is defined as,
\begin{equation}
\bar{\alpha}=\int_0^t H(\alpha \dot{\alpha})\abs{\dot{\alpha}} \dd \tau,
\end{equation}   
with $\tau$ being the pseudo time for integration and $H$ the Heaviside function for disabling damage accumulation during unloading. The history variable $\alpha$ is defined as the active part of the elastic strain energy density, i.e. $\alpha=g(d)\psi_0^+$, and $\dot{\alpha}$ is its rate. It remains to define the fatigue degradation function $f(\bar{\alpha})$, which describes how the material resistance to fracture degrades during cyclic damage. An asymptotic function is used, such that
\begin{equation}
f(\bar{\alpha})= \begin{cases}
1 & \mathrm{if \quad} \bar{\alpha}\leq \alpha_T, \\
\left( \frac{2\alpha_T}{\bar{\alpha}+\alpha_T}\right)^2 & \mathrm{if \quad} \bar{\alpha}\geq \alpha_T,
\end{cases}
\end{equation}
where $\alpha_T = G_\text{c}/(12l)$ is the fatigue crack threshold, only above which fatigue damage can start to accumulate \cite{Carrara2020}.

\section{Methods: Numerical implementation}
\label{sec:FEimplementation}
This section describes the numerical implementation of the above presented electrochemo-mechanical theory for fatigue cracking in electrode particles. For this purpuse, the finite element method (FEM) is used.\\

First, we formulate the governing equations in their weak form, and thus re-write \cref{eq:stress_strong,eq:diffusion_strong,eq:fatigue_PF} as
\begin{subequations}
\begin{gather}\label{eq:varWeakForm}
\int_\Omega \grad(\delta c) \cdot \textbf{J}\dd\Omega-\int_{\Omega}\delta c\derivative{c}{t}\dd \Omega - \int_{\partial\Omega} \delta c \bar{J} \dd S = 0 , \\ 
 \int_\Omega g(d)\bm{\sigma}:\delta\bm{\varepsilon}\dd \Omega - \int_{\partial\Omega}\bar{\textbf{t}}\cdot\delta\textbf{u}\dd S=0,\\
 \int_\Omega\left[-2(1-d) \mathcal{H}\delta d+G_\text{d}\left(\frac{d\delta d}{l}+l\grad{d}\cdot\grad{\delta d}\right)\right]\dd \Omega =0.
\end{gather}
\end{subequations}
Note that the second term on the left hand-side of \cref{eq:fatigue_PF} is eliminated in the weak form. The lithium concentration $c$, the displacement field $\textbf{u}$ and the phase field $d$ are approximated using the following finite element discretisation: 
\begin{equation}
c=\sum_{i=1}^m N_i c_i, \qquad \textbf{u}=\sum_{i=1}^m N_i \textbf{u}_i \quad \text{and} \quad d=\sum_{i=1}^m N_i d_i, 
\end{equation}  
where $m$ is the total number of nodes per element, $N_i$ is the shape function with the subscript $i$ corresponding to the node $i$. The following equations are written in 2D for simplicity but we implement our model in both 2D (plane strain, axisymmetry) and 3D. The gradient terms are approximated using
\begin{subequations}
	\begin{gather}
			[\varepsilon_{11}, \varepsilon_{22}, 2\varepsilon_{12}]^\text{T}=\sum_{i=1}^m \textbf{B}_i\textbf{u}_i, \qquad \grad{c}=\sum_{i=1}^m \textbf{G}_ic_i  \quad \text{and} \quad \grad{d}=\sum_{i=1}^m \textbf{G}_id_i,\\
			\text{with} \quad \textbf{B}_i=\begin{bmatrix}
				\pdv*{N_i}{x} & 0 \\
				0 & \pdv*{N_i}{y} \\
				\pdv*{N_i}{y} & \pdv*{N_i}{x}\\
				\end{bmatrix} \quad \text{and} \quad 
			\textbf{G}_i= [\pdv*{N_i}{x}, \pdv*{N_i}{y}]^\text{T},
	\end{gather}
\end{subequations} 
and the components of the stress tensor in \cref{eq:stress} are  
\begin{equation}
	[\sigma_{11},\sigma_{22},\sigma_{12}]^\text{T}=\sum_{i=1}^m\textbf{C}_0 \left[ \textbf{B}_i\textbf{u}_i - N_ic_i\frac{\Omega}{3}\textbf{I}^c\right],
\end{equation}
where $\textbf{C}_0$ is the linear elastic stiffness matrix and $\textbf{I}^c=[1,1,0]^\text{T}$. Considering that the weak form \cref{eq:varWeakForm} must be ensured for arbitrary $\delta c$, $\delta \textbf{u}$ and $\delta d$, the discrete equation can be expressed with the following residual vectors,  
\begin{subequations}\label{eq:residuals}
    \begin{gather}
        \textbf{r}_i^u=\int_\Omega g(d)\textbf{B}_i^\text{T}\bm{\sigma}\dd \Omega - \int_{\partial\Omega} N_i \bar{\textbf{t}} \dd S, \\
        r_i^d= \int_\Omega\left[-2(1-d)N_i \mathcal{H} + G_\text{d}\left(\frac{N_id}{l}+l\textbf{G}^\text{T}_i\grad d\right)\right]\dd \Omega, \\
        r_i^c=\int_\Omega \left( \textbf{G}_i^\text{T}D\grad c -  \textbf{G}_i^\text{T}\frac{D\Omega_\text{c}c}{RT}\grad\sigma_h  \right)\dd \Omega + \int_{\Omega} N_i \derivative{c}{t} \dd \Omega+ \int_{\partial\Omega} N_i \bar{J} \dd S,
    \end{gather}
\end{subequations}
and the linearised finite element system readily follows; in \cref{Sec:LinearEquation}, we provide explicit expressions for the linearised system and the components of the consistent stiffness matrix. The Newton-Raphson method is used to obtain the solution for $[\textbf{r}_i^u, r_i^d, r_i^c]=[\textbf{0}, 0, 0]$. The governing equations in \cref{eq:varWeakForm} are solved in a monolithic fully coupled manner, rather than using staggered schemes. This is of utmost importance in fatigue problems, as staggered approaches require a significant number of load increments to capture the equilibrium solution \cite{Kristensen2020}. 

The contribution of the gradient of the hydrostatic stress $\sigma_h$ to the diffusivity stiffness matrix is achieved in a decoupled way, by extrapolating from values at integration points and subsequently multiplying $\textbf{G}$ to compute $\grad \sigma_h$ \cite{Martinez-Paneda2018}. The history variables $\mathcal{H}$ and $\bar{\alpha}$ are updated using the following equations
\begin{equation}\label{eq:historical}
	\begin{cases}
		\mathcal{H}^n=\psi^+_0 & \text{if } \psi^+_0>\mathcal{H}^{n-1},\\
		\mathcal{H}^n=\mathcal{H}^{n-1} & \text{if } \psi^+_0 \le \mathcal{H}^{n-1},\\
	\end{cases} \quad \text{and} \quad
	\begin{cases}
		\bar{\alpha}^n=\bar{\alpha}^{n-1}+\abs{\alpha^{n}-\alpha^{n-1}} & \text{if } \alpha^{n}>\alpha^{n-1},\\
		\bar{\alpha}^n=\bar{\alpha}^{n-1} & \text{if } \alpha^{n}\le\alpha^{n-1},\\
	\end{cases}
\end{equation}
where the superscripts $n$ and $n-1$ correspond to the $n$th time step and the previous one respectively. The system linear equation is solved using the PARDISO solver and an implicit Backward Differentiation Formula (BDF). The commercial finite element package \texttt{COMSOL Multiphysics} is used to implement the coupled deformation-diffusion-fracture (fatigue) model, combining the Structural Mechanics, Transport of Diluted Species and Helmholtz Equation modules\footnote{The \texttt{COMSOL} model developed is made freely available at \url{www.empaneda.com/codes}.}.

\section{Numerical experiments}
\label{Sec:FEMresults}

The modelling capabilities of our theory are demonstrated by investigating fatigue cracks in electrode particles. Three case studies are considered, involving idealised particles (cylindrical and spherical) as well as realistic microstructures, which are obtained from advanced imaging techniques. The surface of the electrode particles is subjected to a flux loading $s(t)\bar{J}$, where $s(t)$ is the modified sign function, which is used to switch between charge and discharge. A smooth transition between the two states is assumed to facilitate numerical convergence, occupying only 0.1\% of each one cycle period. The magnitude of the flux $\bar{J}$ [$\mathrm{mol \cdot m^{-2}s^{-1}}$] is defined as 
\begin{equation}\label{eq:flux}
	\bar{J}=c_\text{max}\frac{\text{volume}}{\text{area}}\frac{C}{3600},
\end{equation}  
where $C$ is the C-rate, i.e. the inverse of the time (in hours) to fully charge/discharge a battery. Once the lithium concentration reaches the cut-off values of $c=0$ or the maximum concentration $c=c_\text{max}$, the lithium concentration is then held at these limits. The material properties of the electrode particles are listed in \cref{tb:MaterialProperties}. Among these parameters, increasing Young's modulus and lithium partial molar volume typically leads to an increase in the stress level, while the opposite effect is observed for the diffusion coefficient. Details on the interplay between electrode particle stresses and these parameters can be found in Ref. \cite{Purkayastha2012} and are not explored here. Unless otherwise stated, the characteristic length of the phase field is chosen to be $l=0.01$ \textmu m, following \cite{Klinsmann2016,Klinsmann2016a}. Characterising electrode particle behaviour is challenging but improved insight has been recently gained through nanoindentation \cite{Stallard2022MechanicalBatteries} and image-based modelling techniques \cite{boyce2022}. The initial crack is introduced by applying a non-zero history field $\mathcal{H}$ to the local crack partition domain at $t=0$, using
\begin{equation}\label{eq:initial_crack}
    \mathcal{H}=\alpha_0 \exp\left(-\frac{100s^2}{l^2}\right),
\end{equation}
where $\alpha_0 = 10^{12} \, \mathrm{J/m}^3$ and $s$ is the distance to the crack plane. All our numerical experiments are conducted under isothermal conditions and at room temperature, $T=298$ K. We define a normalised crack domain $\bar{a}_\text{c}$ as the ratio between the region with $d>0.95$ and the domain area (or volume) $\Omega$:
\begin{equation}\label{eq:a_c}
    \bar{a}_\text{c} = \frac{1}{\Omega}\int_\Omega H(d-0.95) \, \dd \Omega \, .
\end{equation}
In addition, similar to \cite{Klinsmann2016}, we quantify unstable crack growth by defining a threshold for the rate of $\bar{a}_\text{c}$. Thus, we assume that unstable crack growth occurs when $\derivative{\bar{a}_\text{c}}{N}>k_\text{un}$, where $N$ is the number of cycles and $k_\text{un}$ is a threshold quantity, such that the unstable crack growth domain is given by,
\begin{equation}\label{eq:UnstableCrack}
  \bar{a}_\text{un}= \int_N \derivative{\bar{a}_\text{c}}{N}H \left(\derivative{\bar{a}_\text{c}}{N}-k_\text{un} \right) \dd N.  
\end{equation}

\noindent In this work, we assume that $k_\text{un}=0.002 \%$. 

Cycle-by-cycle computations enable predicting fatigue damage for arbitrary choices of material, geometry and loading history. However, this insight comes at a high computational cost. Here, exploiting the good scalability and monolitihic nature of our implementation, we report simulations of 300 cycles in 2D problems (\cref{sec:case1,sec:case2}) and of 140 in the 3D case study (\cref{sec:case3}); these numbers are on the same order of the typical lifespan of lithium-ion batteries \cite{Severson2019}. Our focus is on providing the first modelling results of fatigue crack growth due to cyclic flux loading. Hence, the chemistry of electrode particles is kept simple, e.g. constant diffusivity and current. We choose to neglect the impact of SEI formation on newly created crack faces as the magnitude of flux loading is relatively insensitive at high Coulombic efficiencies (99.99\%) and the SEI layer should not change the structural integrity of electrode particles due to its porous and weak nature. The framework can be readily extended to more complex scenarios, such as concentration dependent diffusivity, variable flux and additional flux on cracked surfaces.

\begin{table}[t!]
	\caption{Material properties for lithium manganese oxide  $\mathrm{LiMn_2O_4}$ \cite{Klinsmann2016a}.}
	\label{tb:MaterialProperties}
	\centering
	\begin{tabular}{cccc}
		\hline
	Description	& Symbol & Value & Unit \\
			\hline
	Young's modulus	& $E$ & 93 & GPa   \\
	Poisson's ratio	& $\nu$ & 0.3 & 1    \\
	Critical energy release rate &  $G_\text{c}$	& 10 & $\mathrm{J/m^2}$   \\
	Maximum lithium concentration	& $c_\text{max}$ & 22,900  &  $\mathrm{mol/m^3}$ \\
	Lithium diffusion coefficient & $D$ & $7.08\times10^{-15}$ & $\mathrm{m^2/s}$\\
	Partial molar volume of lithium & $\Omega_\text{c}$ & $3.497\times10^{-6}$ &$\mathrm{m^3/mol}$\\ 
	\hline
	\end{tabular}
\end{table}

\subsection{Cylindrical electrode particle}
\label{sec:case1}

\begin{figure}[t!]
	\centering
	\includegraphics[width=0.9\linewidth]{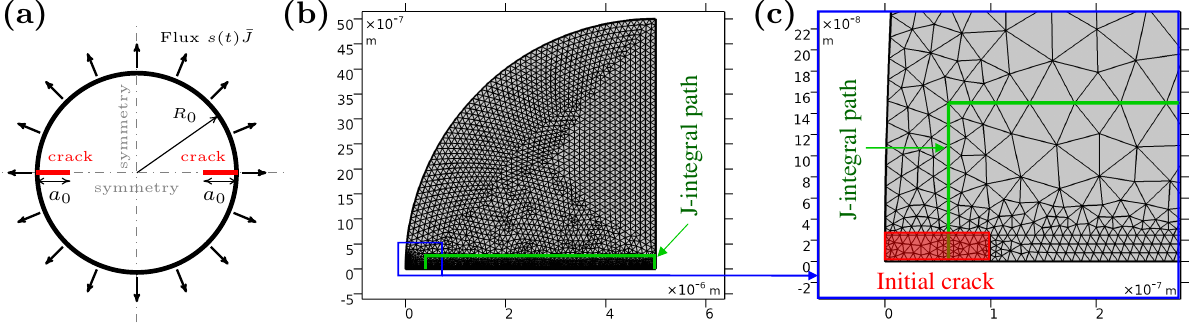}
	\caption{Configuration of a cylindrical particle with two cracks: (a) sketch; (b) computational model (one quarter of the body with symmetric boundary conditions)  and (c) zoom view to showcase the initial crack and the J-integral path.}
	\label{fig:case1_sketch}
\end{figure}

The first example involves a cylindrical particle with two surface cracks, as shown in \cref{fig:case1_sketch}a. 2D, plane strain modelling conditions are adopted, with only one quarter of the particle being simulated using appropriate symmetric boundary conditions at the two straight edges. The finite element mesh, shown in Figs. \ref{fig:case1_sketch}b and \ref{fig:case1_sketch}c, contains 7,875 linear triangular elements. A mesh sensitivity analysis reveals only a 2\% deviation in the calculation of the energy release rate (ERR) from the result obtained using 34,208 elements. The ERR is computed using the J-integral \cite{J.R.Rice1968} over a path surrounding the crack tip, e.g. the green path in \cref{fig:case1_sketch}b; further details of the calculation of the J-integral are given in \cref{Sec:J-integral}. This electrode particle has been studied by Klinsmann et al. \cite{Klinsmann2016,Klinsmann2016a} for a single discharge or charge cycle without considering fatigue damage. 

\begin{figure}[t!]
	\centering
	\includegraphics[width=0.9\linewidth]{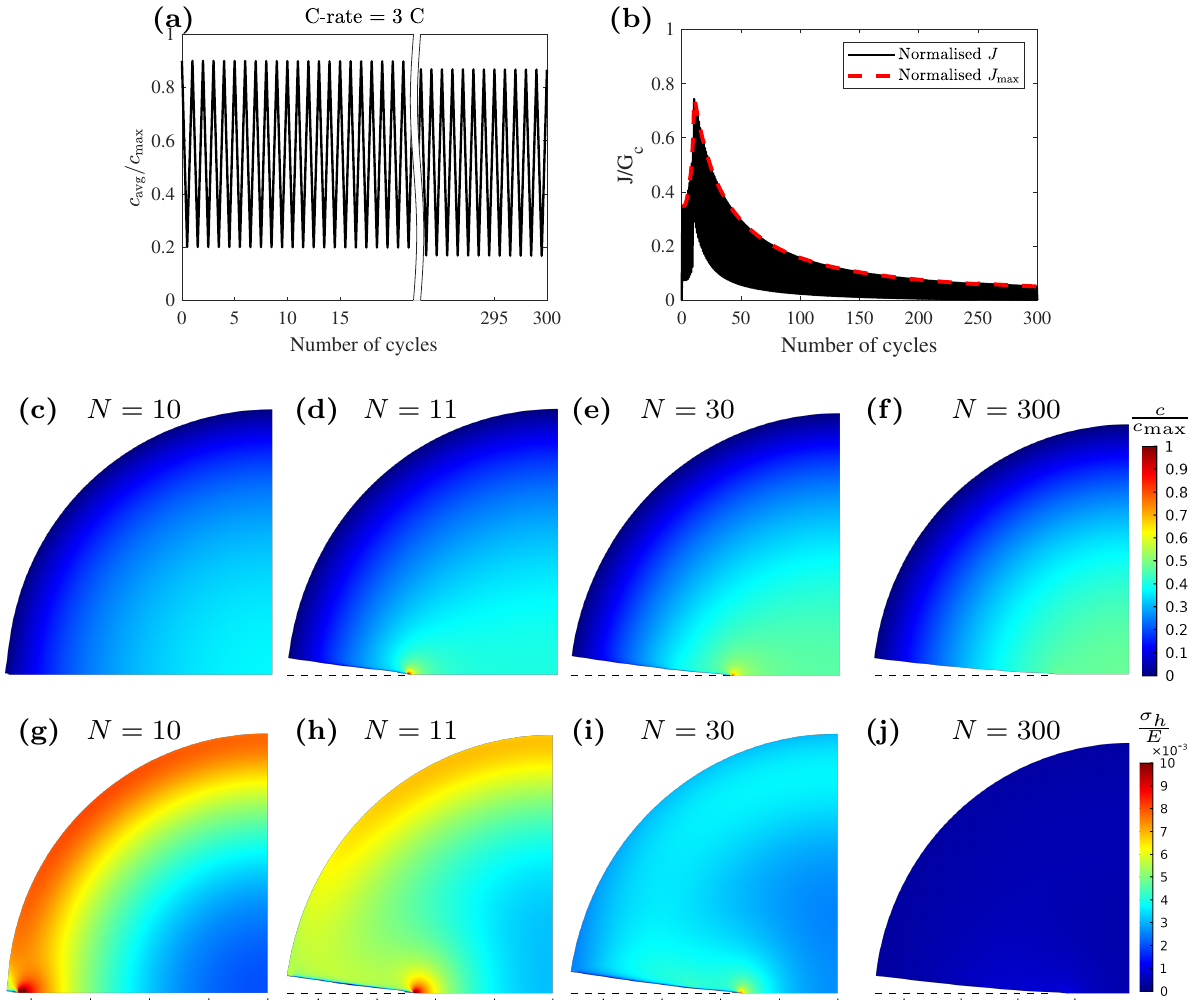}
	\caption{The cycling responses of the cylindrical particle ($R_0=5$  \textmu m) under 3 C current: (a) averaged lithium concentration; (b) energy release rate; and distributions of (c-f) lithium concentration, and (g-j) hydrostatic stress, at the end of discharge for the cycle numbers $N= 10, 11, 30 \text{ and } 300$. The cracked domain $d>0.95$ is removed and the deformed shape is scaled by a factor of 5.}
	\label{fig:case1_3C}
\end{figure}

A typical cycling response for a cylindrical electrode particle under 3 C charge and discharge is shown in \cref{fig:case1_3C}. Unless otherwise stated, the geometry of the particle is characterised by a radius $R_0=5$ \textmu m and an initial crack length equal to $a_\text{0}/R_0=0.02$. A 5 \textmu m particle radius is chosen as a representative, intermediate value within the typical range of electrode particle radii (e.g., between 1 and 15 \textmu m for NMC particles \cite{chen2020a}). As shown in \cref{fig:case1_3C}a, the operating window of the state of charge (SOC) lies between 0.2 and 0.9. The ERR estimated over 300 cycles is presented in \cref{fig:case1_3C}b. The ERR initially increases with the number of cycles, accompanied by small stable crack growth, until cracking becomes unstable after 10 cycles. As justified below, the unstable crack arrests upon approaching the centre of the particle, and then cracking proceeds in a stable manner. This is a different behaviour from the pure mechanical fatigue crack propagation with an accelerating growth rate described by Paris' law.

The distributions of the lithium concentration and hydrostatic stress (at the end of discharge) are depicted in \cref{fig:case1_3C} (e-f) and (g-j) respectively, where it is observed that a high lithium concentration accumulates close to the crack tip. This is because high hydrostatic stresses drive the diffusion of lithium near the crack tip, as indicated by \cref{eq:diffusion_strong}. Three groups of simulations have been used to study the influence of the C-rate, the particle size and the initial crack length on the behaviour of fatigue cracks in electrode particles.\\

\begin{figure}[t!]
	\centering
	\includegraphics[width=1\linewidth]{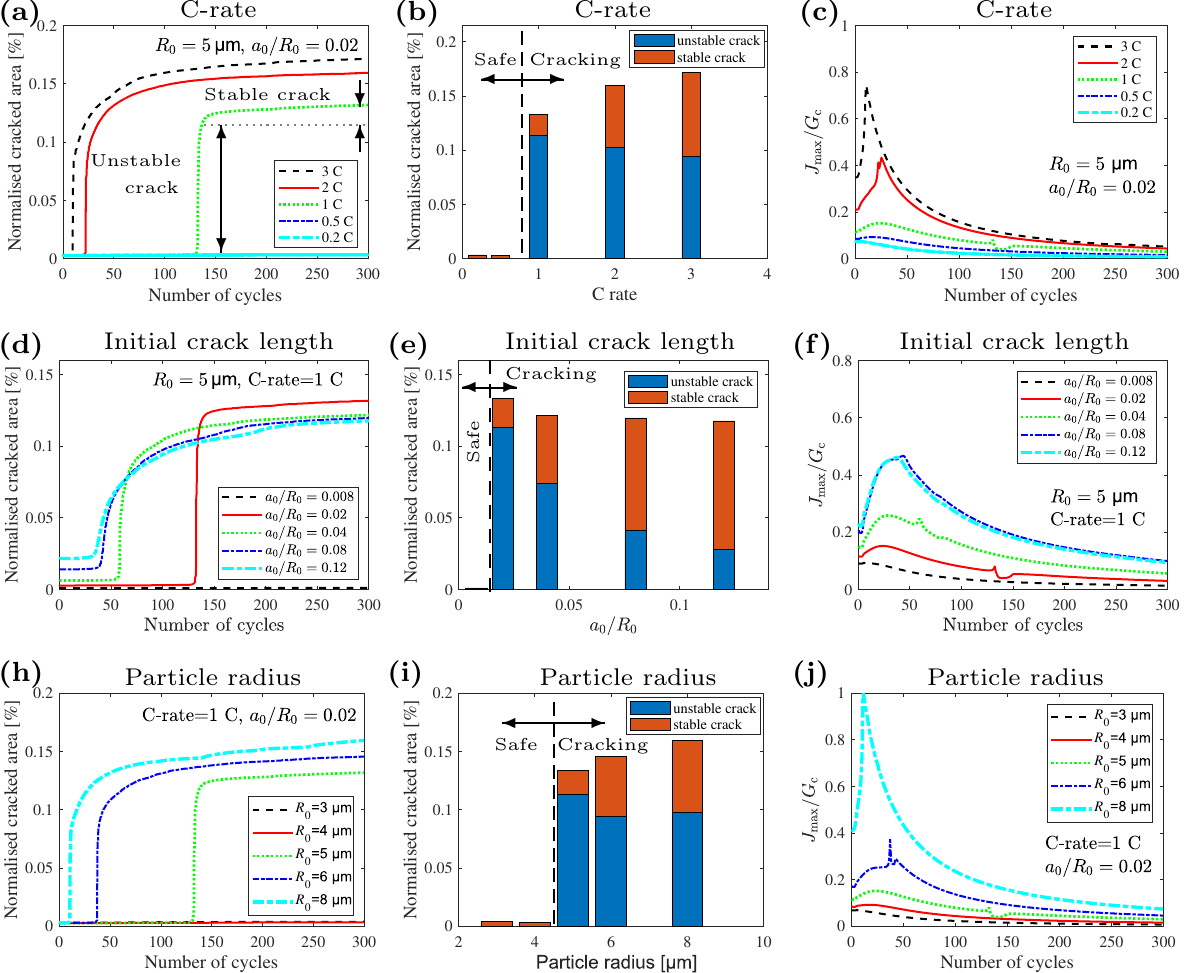}
	\caption{Influence of (a-c) C-rate, (d-f) initial crack length and (g-i) particle radius on fatigue cracking in cylindrical particles: (a, d, g) crack growth versus number of cycles; degree of stable and unstable cracking as a function of (b) C-rate, (e) initial crack length and (h) particle radius, and (c, f, i) normalised magnitude of the maximum energy release rate $J_\text{max}/G_c$ versus number of cycles.}
	\label{fig:case1_result}
\end{figure}

\noindent \textbf{C-rate.} Five C-rates are considered, with $C = 0.2$, 0.5, 1, 2 and 3 for a cylindrical particle with $R_0=5$ \textmu m and $a_\text{0}/R_0=0.02$. The crack growth and the maximum ERR per cycle $J_\text{max}$ are given in \cref{fig:case1_result} (a-c). \cref{fig:case1_result}a shows the percentage of normalised crack area $\bar{a}_c$, with $\bar{a}_c$ estimated using \cref{eq:a_c}. The crack behaviour exhibits three distinct regimes. Firstly, small stable crack growth occurs. This is followed by an unstable cracking phase, where the crack propagates instantaneously until reaching the core of the particle. Finally, following the arrest of the unstable crack, fracture takes place again in a stable manner. Several phenomena lie behind this cracking behaviour. First and foremost, recall the interaction between mechanical stresses and the processes of Li extraction and insertion. During delithiation, the particle surface is subjected to tensile stresses but the stress level diminishes gradually towards the core, with the central region of the particle undergoing compression. The opposite takes place during lithiation, with tensile stresses at the core and a compressive stress state at the outer regions. Thus, the unstable fracture event takes place during a single charging cycle, triggered by the outer tensile stresses and leading to crack arrest once the growing crack approaches the compressive region. After crack arrest, subsequent stable crack growth is driven by the tensile stresses arising at the particle core due to Li transport. Thus, another relevant phenomenon is the diffusion of Li towards regions of high hydrostatic stress, such as the crack tip - see Eq. (\ref{eq:diffusion_strong}). The accumulation of Li facilitates cracking and is particularly relevant during the last stage of stable crack growth. A third aspect to take into consideration is the role of fatigue damage. Unlike the static fracture analyses reported in the literature, fatigue damage leads to an heterogeneous toughness distribution $G_d$. For example, this implies that the unstable crack encounters a growing $G_d$ value as it propagates. Also relevant is that while crack growth takes place during delithiation, lithiation cycles contribute to fatigue damage in the particle core. More importantly, incorporating cyclic damage enables quantifying the number of cycles that take place before unstable particle cracking occurs.\\

The timing of the unstable crack growth event is very sensitive to the C-rate, as unstable fracture occurs once the fatigue damage process has degraded the material toughness $G_c$ sufficiently, relative to the size of the initial crack and the magnitude of the surface tensile stresses resulting from delithiation. Only ten cycles are needed for 3 C conditions while for a current of 1 C more than 130 cycles take place before unstable crack growth. The higher the C-rate, the higher the stress magnitude due to a larger concentration gradient. As a result, the critical crack length for stable cracking decreases with increasing C-rate. The degree of stable and unstable cracking is quantified as a function of the C-rate in Fig. \ref{fig:case1_result}b. It can be observed that there is a critical threshold below which the electrode particle exhibits no crack propagation. For the conditions and lifespan considered here, the critical C-rate is determined to be 0.5. One should note that the critical C-rate is sensitive to the choice of material, chemistry and cell design. For example, 
power cells generally have higher critical C-rates than energy cells due to more homogeneous current distribution, and electrode materials with high energy densities (like silicon) often suffer from large volume changes and have lower critical C-rates \cite{Liu2012}. Finally, in \cref{fig:case1_result}c we report the maximum value of the energy release rate at each cycle, normalised by the fracture toughness $G_c$. In agreement with expectations, we observe how the ERR increases with the C-rate, as a result of larger stresses and concentration gradients. It is also seen that unstable cracking is predicted for maximum values of the ERR that are well below the material toughness $G_c$, due to the degrading effect of fatigue damage. However, if the C-rate is sufficiently small, the magnitude of ERR remains low and no unstable cracking is observed within the tested lifespan.\\  

\noindent \textbf{Initial crack length.} Quantifying the role of the initial crack size is important because, among other factors, the calendaring process in battery manufacturing can lead to initial defects in the electrode particles. Five different initial crack lengths have been considered in a cylindrical particle with $R_0=5$ \textmu m subjected to 1 C cycling: $a_\text{0}/R_0=0.008$, 0.02, 0.04,  0.08 and 0.12. The results obtained are shown in \cref{fig:case1_result}(d-f). The first observation, as seen in Figs. \ref{fig:case1_result}d and \ref{fig:case1_result}e, is that there is a threshold crack size below which the degree of crack growth is negligible. For a particle size of $R_0=5$ \textmu m, 300 cycles and a C-rate of 1 C, critical cracks have an initial size of 0.1 \textmu m or larger. This is a much smaller critical size than the one reported if fatigue effects are neglected (see \cite{Klinsmann2016a}). A second observation is that all $a_0$ cases above the critical level accumulate approximately the same level of cracked domain after 300 cycles, see Figs. \ref{fig:case1_result}d and \ref{fig:case1_result}e. However, the degree of unstable cracking increases as $a_0$ becomes smaller. In other words, the most critical scenario is the one where the initial defect size is just above the critical threshold for cracking. The trends observed can be rationalised by inspecting the evolution of the ERR, shown in Fig. \ref{fig:case1_result}f. Larger cracks lead to higher values of ERR and this leads in turn to an earlier initiation of crack growth, as shown in Fig. \ref{fig:case1_result}d. The later onset of crack growth for the smaller $a_0$ ($N \approx 130$ for $a_0/R_0=0.02$) also implies that the degraded toughness $G_d$ ahead of the initial crack is going to be significantly smaller than that of the pristine material. Also, one must keep in mind the competition between the contributions of the crack size and the crack tip stresses to the fracture driving force. For the surface cracks considered here, those of a smaller size are exposed to larger delithiation stresses while larger cracks have their tips located in a region of smaller stress levels.\\

\noindent \textbf{Particle radius.} The magnitude of the stresses resulting from the inhomogeneous swelling and shrinkage of the particle is governed not only by the C-rate but also by the particle radius, through their effect on the concentration gradient. This sensitivity on particle radius has been observed in reduced order models \cite{Ai2020,Christensen2006} and experiments \cite{Liu2012}, where a critical radius of $R_0=150$ nm has been identified for silicon particles, below which no cracking is observed. Here, five values of particle radius have been considered, $R_0$=3, 4, 5, 6 and 8 \textmu m, with a fixed initial crack length $a_\text{0}/R_0=0.02$ and 1 C flux loading. The results, reported in \cref{fig:case1_result}(g-i), show that larger particles exhibit a higher ERR and a larger cracked surface. Likewise, as shown in Fig. \ref{fig:case1_result}g, the number of charging cycles that the particle can undergo before unstable crack growth occurs increases with its size. Also, the results reveal that there is a critical particle size below which no fatigue cracking is observed, for the number of cycles and loading conditions considered. Thus, for a C-rate of 1 C and an initial crack $a_\text{0}/R_0=0.02$, the critical particle radius equals 0.4 \textmu m.   

\subsection{Spherical electrode particle}
\label{sec:case2}

\begin{figure}[t!]
	\centering
	\includegraphics[width=1\linewidth]{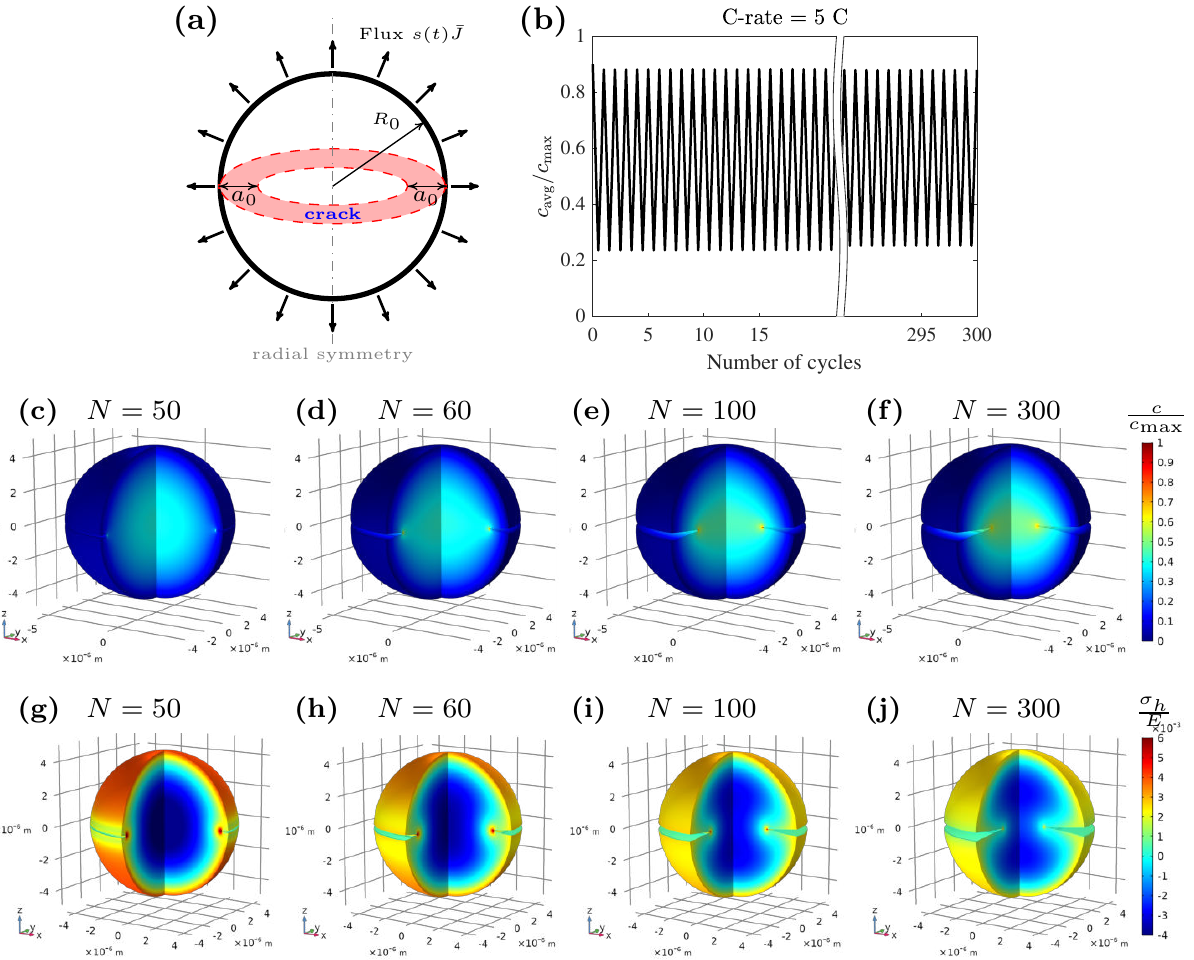}
	\caption{Spherical particle. Geometry, charging conditions and contour plots: (a) particle and crack configuration; (b) average lithium concentration; and distributions of (c-f) lithium concentration, and (g-j) hydrostatic stress, at the end of discharge for the cycle numbers $N=10$, 11, 30 and 300. A quarter of the particle is omitted to showcase the internal domain, the cracked domain $d>0.95$ is removed, and the deformed shape is scaled by a factor of 5. Results are shown for a particle of radius $R_0=5$ \textmu m under 5 C current.}
	\label{fig:case2_sketch}
\end{figure}

Our second set of numerical experiments addresses the fatigue failure of spherical electrode particles. An initial circular crack is assumed, as this is the crack domain that results from two edge cracks in a 3D particle undergoing one discharge (see \citealp{Klinsmann2016a}). The particle and initial crack configurations are shown in \cref{fig:case2_sketch}. Axisymmetric conditions are exploited and as such the mesh resembles the one used in cylindrical particles (Section \ref{sec:case1}), but with the use of axisymmetric finite elements. The averaged concentration for the spherical particle under 5 C cycling is given in \cref{fig:case2_sketch}b with the SOC being between 0.2 and 0.9. The distributions of lithium concentration and hydrostatic stress are presented in \cref{fig:case2_sketch} (c-f) and \cref{fig:case2_sketch} (g-j), respectively. Contours are shown at the end of discharge for the cycle numbers $N=50$, 60, 100 and 300. One quarter of the particle is omitted for the sake of a better visualisation. The results reveal high lithium concentrations and hydrostatic stresses at the crack front, with the hydrostatic stress level decreasing as the crack penetrates towards the centre of the particle.\\    

\begin{figure}[t!]
	\centering
	\includegraphics[width=1\linewidth]{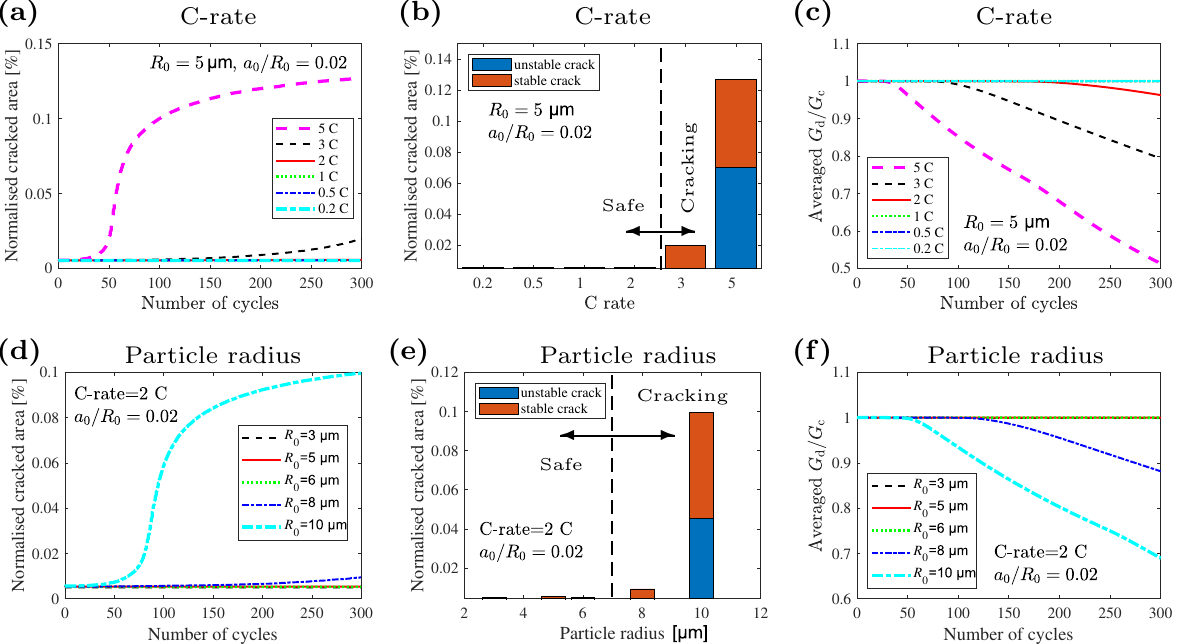}
	\caption{Influence of (a-c) C-rate and (d-f) particle radius on fatigue cracking in spherical particles: (a and d) crack growth versus number of cycles; degree of stable and unstable cracking as a function of (b) C-rate and (e) particle radius, and (c and f) averaged fatigue degradation of fracture toughness over the problem domain.}
	\label{fig:case2_result}
\end{figure}

\noindent\textbf{C-rate.} The influence of the C-rate is investigated by considering the following values: 0.2, 0.5, 1, 2, 3, and 5 C. The spherical particle has a radius of $R_0=5$ \textmu m and the initial crack length equals $a_\text{0}/R_0=0.02$. The results are shown in \cref{fig:case2_result}(a-c) in terms of the change in normalised crack area as a function of the number of cycles, the degree of stable and unstable crack growth, and the averaged degradation toughness $G_d$. Here, the averaged $G_d$ is computed by integrating over the entire domain. The first notable observation is that there is a C-rate threshold below which no cracking is observed. Specifically, for the particle configuration ($R_0=5$ \textmu m, $a_\text{0}/R_0=0.02$) and battery lifespan ($N=300$) considered, charging rates equal or below 2 C show negligible damage. As shown in \cref{fig:case2_result}c, the degraded toughness ($G_d$) equals the undegraded one ($G_c$) over the entire lifespan for the C-rates of 0.2, 0.5 and 1. This implies that the stresses and strains associated with the swelling and shrinkage of the particle are not large enough to reach the fatigue threshold $\alpha_T$. For the 3 C charge a certain degree of stable crack growth is observed, see Figs. \ref{fig:case2_result} (a-b), but the accumulated damage over 300 cycles is not sufficient to trigger the propagation of large cracks. Significant crack propagation is observed for the highest current, 5 C, with unstable cracking occurring after approximately 50 cycles. Compared with the results obtained for a cylindrical particle of the same radius and initial crack, Fig. \ref{fig:case1_result}, we notice the spherical particle can withstand a larger C-rate without exhibiting significant cracking.\\

\noindent\textbf{Particle radius.} We proceed to show the results obtained for selected values of the particle radius: $R_0=$3, 5, 6, 8 and 10 \textmu m. The initial crack length is assumed to be $a_\text{0}/R_0=0.02$ and the spherical particle is subjected to a cycling charging of 2 C. The results obtained are shown in \cref{fig:case2_result}(d-f). In agreement with expectations, and with the results obtained for the cylindrical particle (Fig. \ref{fig:case1_result}), a higher degree of fatigue damage is observed in larger particles. The larger the particle size, the more significant the effect of particle volume changes. In fact, unstable cracking is observed only for the largest particle radius, $R_0=$10 \textmu m, where fast crack growth is attained before reaching 100 cycles. The case with $R_0=$8 \textmu m also appears to show a certain degree of stable crack growth and a more noticeable drop in the degraded toughness $G_d$. For particles with radius equal to or smaller than $R_0=$6 \textmu m the degradation of the material toughness is negligible and thus no cracking is observed. 

\subsection{Realistic electrode particle}
\label{sec:case3}

\begin{figure}[t!]
	\centering
	\includegraphics[width=1\linewidth]{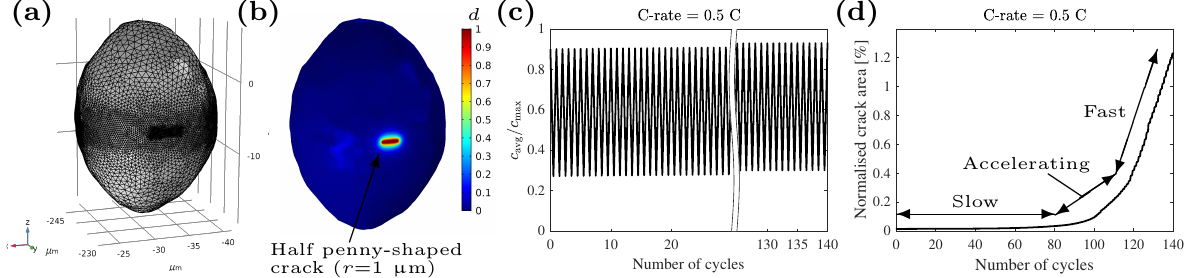}
	\caption{3D realistic electrode particle: (a) mesh; (b) initial defect; (c) averaged lithium concentration during cycling; and (d) normalised cracked area versus number of cycles. The initial defect is introduced at a location (-31.5, -230, -6) using \cref{eq:initial_crack} and with the shape of a half-penny crack.}
	\label{fig:case3_problem}
\end{figure}

Finally, we couple our modelling framework with X-ray CT imaging to predict the fatigue cracking of realistic electrode particles. The particle geometry is taken from the X-ray tomographic microscopy open data set presented by Ebner et al. \cite{Ebner2013b}. As shown in \cref{fig:case3_problem}a, the particle has an ellipsoidal geometry with a longer radius of approximately 15 \textmu m and two smaller radii of approximately 10 \textmu m. The problem domain is discretised with linear tetrahedral elements, using slightly more than 3M degrees-of-freedom. The mesh is sufficiently refined throughout the particle to resolve the phase field length scale, which is taken to be $l=0.6$ \textmu m. The localisation of the damage is facilitated by introducing an heterogeneity in the form of a surface crack, as shown and described in Fig. \ref{fig:case3_problem}. This is also necessary to resemble conditions in commercial cells - during battery manufacturing, electrode materials are calendered to increase the volumetric density, but this process introduces initial cracks in the electrode particles \cite{Lu2020NC}. In addition, we model the role of cracks in blocking Li transport by introducing the following modification to the diffusion coefficient:
\begin{equation}
    D=D\cdot g(d), \quad \text{ when } d>0.95,
\end{equation}

\noindent where the degradation function $g(d)$ is defined in \cref{eq:degradation}. A more comprehensive approach can also be adopted, in which the additional lithium flux resulting from the increased surface area is accounted for (see \cite{Zhao2016b}). The boundary conditions resemble those of the previous case studies; the particle surface is under a cyclic and uniform flux loading, using \cref{eq:flux}. A C-rate of 0.5 is considered and the resulting cyclic evolution of the averaged lithium concentration is given in \cref{fig:case3_problem}c. Relative to idealised particles under the same C-rate, one would expect realistic particles to show a larger degree of damage due to the higher lithium concentration gradients and stresses present.

\begin{figure}[t!]
	\centering
	\includegraphics[width=1\linewidth]{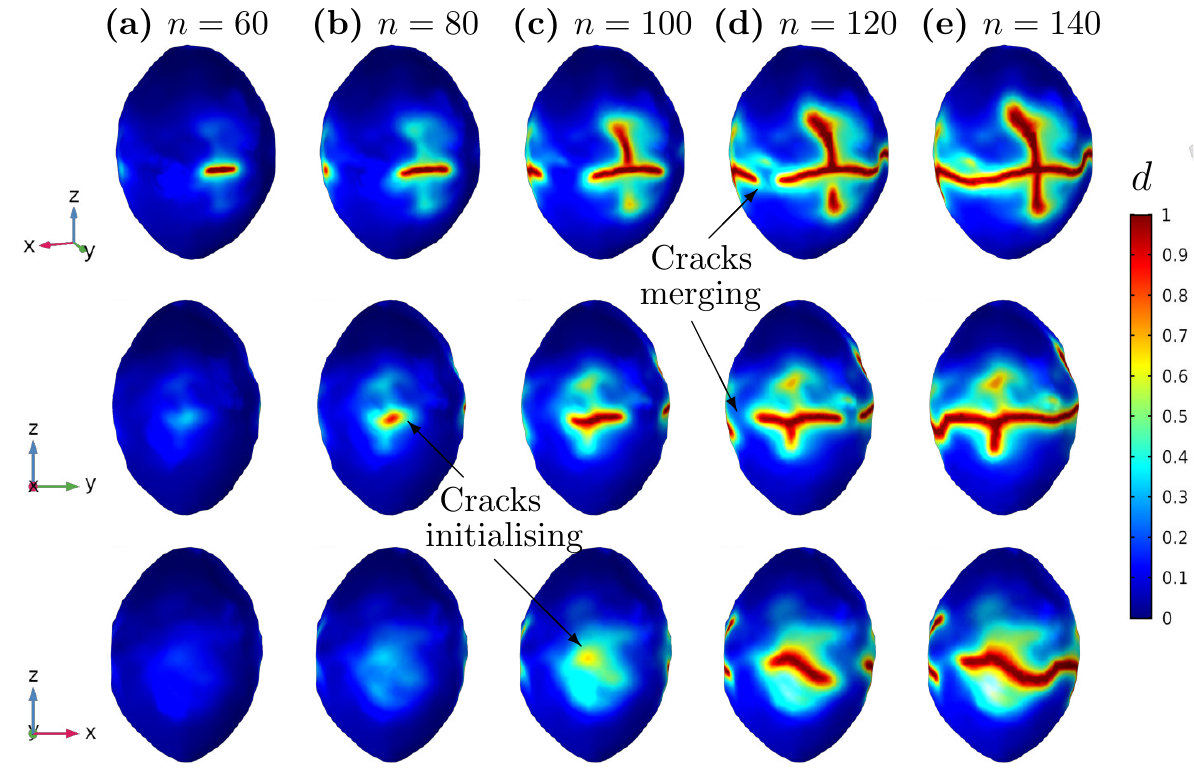}
	\caption{3D realistic electrode particle: evolution of the phase field $d$ contours for the following cycle numbers: (a) $N=60$, (b) $N=80$, (c) $N=100$, (d) $N=120$, and (e) $N=140$. The three rows correspond to perspective views of the 3D particle, the $yz$-plane, and the $xz$-plane, respectively.}
	\label{fig:case3_p}
\end{figure}

\begin{figure}[t!]
	\centering
	\includegraphics[width=1\linewidth]{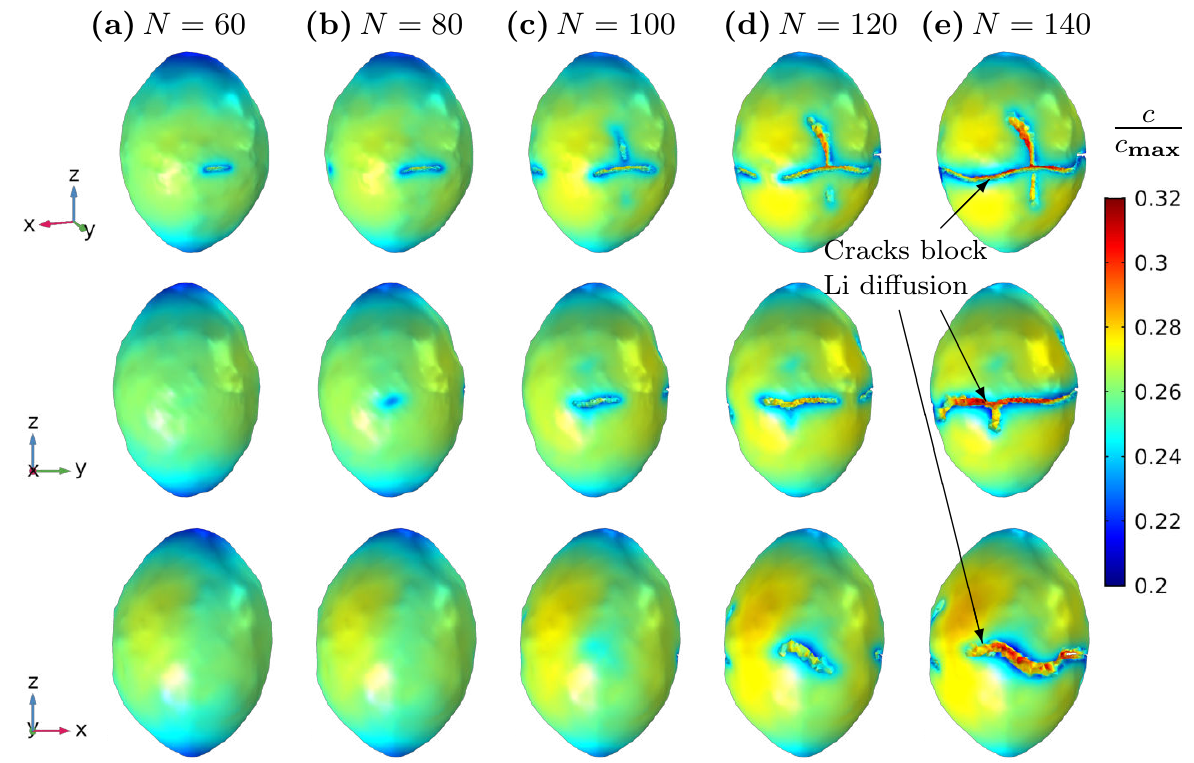}
	\caption{3D realistic electrode particle: evolution of the normalised Li concentration $c/c_{max}$ contours for the following cycle numbers: (a) $N=60$, (b) $N=80$, (c) $N=100$, (d) $N=120$, and (e) $N=140$. The three rows correspond to perspective views of the 3D particle, the $yz$-plane, and the $xz$-plane, respectively. Elements with $d>0.95$ have been removed.}
	\label{fig:case3_c}
\end{figure}

The fatigue crack growth behaviour is shown in \cref{fig:case3_problem}d, in terms of cracked domain versus number of cycles, and in Fig. \ref{fig:case3_p}, in terms of phase field contours for different loading stages. As it can be seen in \cref{fig:case3_problem}d, the averaged evolution of the cracked area exhibits three distinct stages. First, the cracked domain increases in a slow manner. This is followed by an acceleration stage, where fatigue crack growth rates increase noticeably. Subsequently, the cracked domain experiences a regime of fast growth, leading to significant particle cracking. This integrated, macroscopic picture is the result of complex crack interactions, as shown in Fig. \ref{fig:case3_p}. The first stage, lasting up to approximately 80 cycles, is characterised by a slow propagation of the initial crack. Then two cracks nucleate on the sides of the initial crack, accelerating the growth of the cracked area, as shown in Figs. \ref{fig:case3_p} (b-c). The normalised crack area increases fast in the last stage, when the cracks merge together in \cref{fig:case3_p} (d-e), indicating that the electrode particle is approaching the end of its lifespan. Based on the behaviour observed for the idealised particles, we speculate that the fast crack propagation stage could be followed by a stable crack growth regime, once all cracks have merged into one and grow inwards, towards the less stressed particle regions. Thus, the analysis of a realistic particle microstructure reveals the existence a regime of exponential increase in crack area with cycle number, which could be approximated with Paris' law. Finally, the evolution of the lithium concentration distribution as a function of the number of cycles is shown in Fig. \ref{fig:case3_c}, where the areas with $d>0.95$ have been removed to represent the location of cracks. It can be observed that Li accumulates at the crack fronts and, mainly, in regions where the presence of cracks blocks the diffusion path. Overall, we observe a significant influence of particle geometry in both cracking and diffusion behaviour. Most notably, the consideration of a realistic microstructure shows the nucleation of cracks in concave regions, see \cref{fig:case3_p} (b-c), which results in a rapid increase of the cracked area. Hence, the results demonstrate the importance of particle microstructure on the cyclic stability of electrode materials, which is seldom studied in reduced order models, e.g. the Newman's battery model \cite{DFN1993}.

\subsection{Analysis of the results and discussion}

Simulations of cylindrical, spherical and realistic electrode particles under multiple cycles of charge and discharge were conducted to study their fatigue resistance. The influence of particle shape and size, C-rate and initial crack length was assessed and their critical values, below which no damage occurs, were quantified. In agreement with expectations, we found that fatigue cracking is enhanced with higher values of particle radius and C-rate, as these lead to larger concentration gradients and thus volume change-induced stresses. The role of the initial crack length is two-fold. No crack growth is observed if the initial defect size is sufficiently short but once the critical threshold is surpassed, shorter cracks lead to a higher degree of unstable crack growth. The trends observed are a result of the contributions of chemical strains, fatigue damage accumulation and Li transport ahead of crack tips. Previous studies where limited to one half-cycle and the use of static fracture formulations; we shall discuss the implications of incorporating fatigue damage. \cref{tb:summary} summarises the critical values of C-rate, particle radius and initial crack length obtained in the analysis of cylindrical and spherical particles. The results are compared with those obtained by Klinsman et al. \cite{Klinsmann2016a} for one discharge process using a phase field model for static, monotonic fracture. Substantial differences are observed; when fatigue damage is taken into account, the reduction in critical C-rate, particle radius and initial crack size is of 90\%, 20\%, and 60\%, respectively. Thus, it is imperative to take fatigue cracks into account in the design of battery control systems. The critical C-rates reported can be used for risk assessment of particle cracking and to guide electrode manufacturing (e.g., favouring the use of spherical shapes). However, one should note that the predictions obtained are specific to the material properties and conditions considered. Other factors that can influence these predictions include the role of the particle microstructure, non-linear diffusion, and heterogeneous electrochemical and fracture behaviour within the electrode \cite{boyce2022}.


\begin{table}[t!]
	\caption{Summary of the critical values for crack propagation in electrode particles computed under cyclic flux. The results are compared with the values obtained for a single discharge by Klinsman et al. \cite{Klinsmann2016a} under otherwise equivalent conditions.}\label{tb:summary}
	\begin{tabular}{ccccc}
		\hline
		 & \multicolumn{2}{c}{Cylindrical particle} & \multicolumn{2}{c}{Spherical particle}\\
		 & Cycling &  Discharge  & Cycling & Discharge \\
		\cline{2-5}
		Critical C-rate & 0.5 C & 5 C & 2 C & 5 C\\
		Critical particle radius [\textmu m] & 4	& 5 & 6 & - \\
		Critical initial crack length $a_\text{0}/R_0$& 0.008 & 0.02 & - & - \\
		\hline
	\end{tabular}
\end{table}

The modelling of fatigue cracking in idealised and realistic particle geometries reveals a four stage process. First, crack growth takes place at a very slow rate. Eventually the speed of crack growth accelerates, and this is soon followed by fast, unstable cracking. Finally, the crack arrests as it reaches regions of low stress, and later resumes its growth in a stable manner, driven by fatigue damage and Li accumulation at crack tips. While the use of Paris' law might provide a good approximation to certain stages of the fatigue crack behaviour observed, whole-life predictions require a more comprehensive approach. Also, our results show that complex cracking phenomena, such as the interaction between multiple cracks, take place early in the fracture of realistic particles. Thus, the use of semi-empirical models that are based on Paris' law and a number of idealisations, such as equally-spaced short cracks \cite{Deshpande2012}, is deemed unsuitable for the analysis and design of fatigue-resistant electrodes. The results shown emphasise the importance of using physically-motivated models and the combination of these models with advanced imaging techniques.

\section{Conclusions}
\label{Sec:ConcludingRemarks}

We have presented a novel electrochemo-mechanical framework for modelling fatigue cracking in battery electrode particles. The model combines a stress-driven extended version of Fick's law for lithium diffusion, intercalation induced volumetric strains, a phase field description of cracks, and a fatigue degradation scheme. We used our framework to gain new insight into the fatigue behaviour of cylindrical, circular and realistic particle geometries undergoing hundreds of charge/discharge cycles. The evolution of the cracked domain was predicted as a function of the number of cycles and the results were interpreted in terms of the contributions from Li transport, chemical strains and fatigue damage mechanics. Critical values of the C-rate, particle size and initial crack length are reported, below which no fatigue damage is observed. Main findings include:

\begin{itemize}
 \item The susceptibility to fatigue damage increases with increasing C-rates and particle sizes. However, a larger degree of unstable cracking is observed by reducing the initial crack length, provided its size is above the critical value.
 \item The critical values of C-rate, particle size and initial crack length are significantly smaller than those reported in the literature, based on a half-cycle analysis without accounting for fatigue damage.
 \item The growth of the cracked domain as a function of the number of cycles exhibits four regimes: slow growth, accelerating growth, fast unstable growth and, after crack arrest, Li diffusion-driven stable crack growth. This behaviour cannot be captured by simple, semi-empirical models.
 \item The coupling with X-CT image analysis enables simulating realistic particle geometries, revealing a fatigue behaviour governed by complex phenomena such as multiple crack interactions.
\end{itemize}

The theoretical and computational framework presented provides a platform for predicting the role of crack mechanics on battery performance, from assessing the viability of design strategies against particle cracking to gaining insight into the regime of fast capacity degradation. Future work will aim at linking particle level analyses with electrode level behaviour, exploiting recent progress in image-based modelling techniques to gain insight into the mechanisms of electrode fragmentation and enable the design of next-generation electrodes.

\section*{Acknowledgements}
\label{Sec:Acknowledgeoffunding}
This work was kindly supported by the EPSRC Faraday Institution Multi-Scale Modelling project (EP/S003053/1, grant number FIRG003). W.Ai additionally acknowledges financial support from Jiangsu Key Laboratory of Engineering Mechanics, Southeast University and from the Fundamental Research Funds for the Central Universities grant number (4060692201/016). E. Mart\'{\i}nez-Pa\~neda additionally acknowledges financial support from UKRI's Future Leaders Fellowship programme [grant MR/V024124/1].

\section*{CRediT author contribution statement}
\label{Sec:CRediT}
\noindent \textbf{Weilong Ai}: Methodology, Software, Investigation, Data Curation, Formal analysis, Writing - Original Draft. \textbf{Billy Wu}: Project administration, Funding acquisition, Writing - Review \& Editing. \textbf{Emilio Mart\'{\i}nez-Pa\~neda}: Conceptualization, Software, Project administration, Funding acquisition, Writing - Review \& Editing.

\section*{Declaration of competing interest}
The authors declare no competing financial interests.

\appendix
\renewcommand{\thefigure}{A.\arabic{figure}}
\setcounter{figure}{0}
\section{Path dependence of the J-integral}
\label{Sec:J-integral}

In an electrochemical system with lithium diffusion, the free energy in a solid domain contains not only the elastic strain energy but also the energy contributions resulting from solute diffusion. Accordingly, in a 2D setting, the J-integral is reformulated as follows: \cite{J.R.Rice1968,Gao2013a}
\begin{equation}\label{eq:J_integral}
	J=\int_\Gamma \left(W\dd y-\sigma_{ij} n_j u_{i,1} \dd s\right) +\int_A \Omega_\text{c} \sigma_{kk} \pdv{c}{x} \dd A,
\end{equation}

\noindent where $W$ is the elastic strain energy density, $\Gamma$ is the integration path, $n_j$ is the component of the normal vector of the path $\Gamma$ and $A$ is the area surrounded by the path $\Gamma$. The path independence of the modified J-integral in \cref{eq:J_integral} is assessed by investigating a cylindrical electrode particle with two cracks under 5 C discharge. The J-integral measurements obtained from three integration paths are almost identical throughout the particle lifespan, with the largest differences being below 4\%.

\renewcommand{\thefigure}{B.\arabic{figure}}
\setcounter{figure}{0}
\section{Additional details of numerical implementation}
\label{Sec:LinearEquation}

Here, we provide explicit expressions for the components of the the linearised system of equations. First, the consistent stiffness matrices are obtained by differentiating the residuals in \cref{eq:residuals} with respect to the nodal variables,
\begin{subequations}\label{eq:LinearEquation}
	\begin{gather}
		\textbf{K}^u_{ij} =\derivative{\textbf{r}^u_i}{\textbf{u}_j} = \int_\Omega g(d)\textbf{B}_i^\text{T}\textbf{C}_0\textbf{B}_j \dd V, \\
		 \textbf{K}^c_{ij}=\derivative{r^c_i}{c_j} = \int_\Omega\left( \textbf{G}_i^\text{T}D\textbf{G}_j-\textbf{G}_i^\text{T} \frac{D\Omega_\text{c}}{RT}N_j\textbf{G}_j \sigma_h\right)\dd \Omega, \\
	\textbf{K}^d_{ij}= \derivative{r^d_i}{d_j}= \int_\Omega\left[ \left(2\mathcal{H}+\frac{G_\text{d}}{l}\right)N_iN_j+G_\text{d}l\textbf{G}^\text{T}_i\textbf{G}_j\right]\dd V,\\
	\textbf{K}^{uc}_{ij} =\derivative{\textbf{r}^u_i}{\textbf{c}_j} = - \int_\Omega g(d)\textbf{B}_i^\text{T}\textbf{C}_0\textbf{I}^{c}N_j\frac{\Omega}{3} \dd V,
	\end{gather}
\end{subequations} 
where the subscripts $i$ and $j$ for the bold variables correspond to the contribution from the nodes $i$ and $j$ rather than the components of the matrix. $\textbf{K}^{cu}_{ij}=0$ is assumed, following the decoupled approach for the gradient of the hydrostatic stress outlined in Section \ref{sec:FEimplementation} and in \cite{Martinez-Paneda2018}. The linearised finite element system then reads:
\begin{equation}
	\begin{bmatrix}
		\textbf{K}^u & \textbf{0} & \textbf{K}^{uc} \\
		\textbf{0} &		\textbf{K}^d &  \textbf{0} \\
						 \textbf{K}^{cu} & \textbf{0} & \textbf{K}^c
	\end{bmatrix}\begin{bmatrix}
	\textbf{u} \\ \textbf{d} \\ \textbf{c}\end{bmatrix}+\begin{bmatrix}
				\textbf{0} & \textbf{0} & \textbf{0} \\
		\textbf{0} &		\textbf{0} &  \textbf{0} \\
		\textbf{0} & \textbf{0} & \textbf{M}
	\end{bmatrix}\begin{bmatrix}
	\dot{\textbf{u}} \\ \dot{\textbf{d}} \\ \dot{\textbf{c}}\end{bmatrix}+\begin{bmatrix}
\textbf{f}^u \\ \textbf{0} \\ \textbf{f}^c\end{bmatrix}=\begin{bmatrix}
		\textbf{r}^u \\ \textbf{r}^d \\ \textbf{r}^c
\end{bmatrix},
\end{equation}

where the remaining terms are given by,
\begin{equation}\label{eq:Forces}
		 \textbf{M}_{ij}=\int_\Omega N_i N_j \dd \Omega, \quad \textbf{f}^u_i=-\int_{\partial\Omega} N_i \bar{\textbf{t}} \dd S, \quad
		  \textbf{f}^c_i=  \int_{\partial\Omega} N_i \bar{J} \dd S.
\end{equation} 



\bibliographystyle{elsarticle-num}
\bibliography{references}

\begin{thebibliography}{10}
\expandafter\ifx\csname url\endcsname\relax
  \def\url#1{\texttt{#1}}\fi
\expandafter\ifx\csname urlprefix\endcsname\relax\def\urlprefix{URL }\fi
\expandafter\ifx\csname href\endcsname\relax
  \def\href#1#2{#2} \def\path#1{#1}\fi

\bibitem{Palacin2016}
M.~R. Palac{\'{i}}n, A.~de~Guibert,
  \href{https://www.science.org/doi/10.1126/science.1253292}{{Why do batteries
  fail?}}, Science 351~(6273) (2016) 1253292.
\newblock \href {https://doi.org/10.1126/science.1253292}
  {\path{doi:10.1126/science.1253292}}.

\bibitem{Lin2017}
N.~Lin, Z.~Jia, Z.~Wang, H.~Zhao, G.~Ai, X.~Song, Y.~Bai, V.~Battaglia, C.~Sun,
  J.~Qiao, K.~Wu, G.~Liu, {Understanding the crack formation of graphite
  particles in cycled commercial lithium-ion batteries by focused ion beam -
  scanning electron microscopy}, Journal of Power Sources 365 (2017) 235--239.
\newblock \href {https://doi.org/10.1016/j.jpowsour.2017.08.045}
  {\path{doi:10.1016/j.jpowsour.2017.08.045}}.

\bibitem{Zheng2012}
H.~Zheng, L.~Zhang, G.~Liu, X.~Song, V.~S. Battaglia,
  \href{http://dx.doi.org/10.1016/j.jpowsour.2012.06.045}{{Correlationship
  between electrode mechanics and long-term cycling performance for graphite
  anode in lithium ion cells}}, Journal of Power Sources 217 (2012) 530--537.
\newblock \href {https://doi.org/10.1016/j.jpowsour.2012.06.045}
  {\path{doi:10.1016/j.jpowsour.2012.06.045}}.

\bibitem{Muller2018}
S.~M{\"{u}}ller, P.~Pietsch, B.~E. Brandt, P.~Baade, V.~De~Andrade,
  F.~De~Carlo, V.~Wood,
  \href{http://dx.doi.org/10.1038/s41467-018-04477-1}{{Quantification and
  modeling of mechanical degradation in lithium-ion batteries based on
  nanoscale imaging}}, Nature Communications 9~(1) (2018) 1--8.
\newblock \href {https://doi.org/10.1038/s41467-018-04477-1}
  {\path{doi:10.1038/s41467-018-04477-1}}.

\bibitem{birkl2017}
C.~R. Birkl, M.~R. Roberts, E.~McTurk, P.~G. Bruce, D.~A. Howey,
  \href{http://dx.doi.org/10.1016/j.jpowsour.2016.12.011}{{Degradation
  diagnostics for lithium ion cells}}, Journal of Power Sources 341 (2017)
  373--386.
\newblock \href {https://doi.org/10.1016/j.jpowsour.2016.12.011}
  {\path{doi:10.1016/j.jpowsour.2016.12.011}}.

\bibitem{Deshpande2012}
R.~Deshpande, M.~Verbrugge, Y.~Cheng, J.~Wang, P.~Liu, {Battery Cycle Life
  Prediction with Coupled Chemical Degradation and Fatigue Mechanics}, Journal
  of The Electrochemical Society 159~(10) (2012) A1730--A1738.
\newblock \href {https://doi.org/10.1149/2.049210jes}
  {\path{doi:10.1149/2.049210jes}}.

\bibitem{Edge2021}
J.~S. Edge, S.~O’Kane, R.~Prosser, N.~D. Kirkaldy, A.~N. Patel, A.~Hales,
  A.~Ghosh, W.~Ai, J.~Chen, J.~Yang, S.~Li, M.~C. Pang, L.~Bravo~Diaz,
  A.~Tomaszewska, M.~W. Marzook, K.~N. Radhakrishnan, H.~Wang, Y.~Patel, B.~Wu,
  G.~J. Offer, {Lithium ion battery degradation: what you need to know},
  Physical Chemistry Chemical Physics 23~(14) (2021) 8200--8221.
\newblock \href {https://doi.org/10.1039/d1cp00359c}
  {\path{doi:10.1039/d1cp00359c}}.

\bibitem{mukhopadhyay2014}
A.~Mukhopadhyay, B.~W. Sheldon,
  \href{http://dx.doi.org/10.1016/j.pmatsci.2014.02.001}{{Deformation and
  stress in electrode materials for Li-ion batteries}}, Progress in Materials
  Science 63 (2014) 58--116.
\newblock \href {https://doi.org/10.1016/j.pmatsci.2014.02.001}
  {\path{doi:10.1016/j.pmatsci.2014.02.001}}.

\bibitem{zhang2017}
S.~Zhang, K.~Zhao, T.~Zhu, J.~Li, {Electrochemomechanical degradation of
  high-capacity battery electrode materials}, Progress in Materials Science 89
  (2017) 479--521.
\newblock \href {https://doi.org/10.1016/j.pmatsci.2017.04.014}
  {\path{doi:10.1016/j.pmatsci.2017.04.014}}.

\bibitem{Zhao2019a}
Y.~Zhao, P.~Stein, Y.~Bai, M.~Al-Siraj, Y.~Yang, B.~X. Xu, {A review on
  modeling of electro-chemo-mechanics in lithium-ion batteries}, Journal of
  Power Sources 413 (2019) 259--283.
\newblock \href {https://doi.org/10.1016/j.jpowsour.2018.12.011}
  {\path{doi:10.1016/j.jpowsour.2018.12.011}}.

\bibitem{Christensen2006}
J.~Christensen, J.~Newman, {Stress generation and fracture in lithium insertion
  materials}, Journal of Solid State Electrochemistry 10~(5) (2006) 293--319.
\newblock \href {https://doi.org/10.1007/s10008-006-0095-1}
  {\path{doi:10.1007/s10008-006-0095-1}}.

\bibitem{Christensen2006a}
J.~Christensen, J.~Newman, {A Mathematical Model of Stress Generation and
  Fracture in Lithium Manganese Oxide}, Journal of The Electrochemical Society
  153~(6) (2006) A1019.
\newblock \href {https://doi.org/10.1149/1.2185287}
  {\path{doi:10.1149/1.2185287}}.

\bibitem{zhang2007}
X.~Zhang, W.~Shyy, A.~Marie~Sastry, {Numerical Simulation of
  Intercalation-Induced Stress in Li-Ion Battery Electrode Particles}, Journal
  of The Electrochemical Society 154~(10) (2007) A910.
\newblock \href {https://doi.org/10.1149/1.2759840}
  {\path{doi:10.1149/1.2759840}}.

\bibitem{li2017}
J.~Li, N.~Lotfi, R.~G. Landers, J.~Park, {A Single Particle Model for
  Lithium-Ion Batteries with Electrolyte and Stress-Enhanced Diffusion
  Physics}, Journal of The Electrochemical Society 164~(4) (2017) A874--A883.
\newblock \href {https://doi.org/10.1149/2.1541704jes}
  {\path{doi:10.1149/2.1541704jes}}.

\bibitem{Purkayastha2012}
R.~Purkayastha, R.~M. McMeeking, {A linearized model for lithium ion batteries
  and maps for their performance and failure}, Journal of Applied Mechanics,
  Transactions ASME 79~(3) (2012).
\newblock \href {https://doi.org/10.1115/1.4005962}
  {\path{doi:10.1115/1.4005962}}.

\bibitem{rieger2016c}
B.~Rieger, S.~V. Erhard, K.~Rumpf, A.~Jossen, {A New Method to Model the
  Thickness Change of a Commercial Pouch Cell during Discharge}, Journal of The
  Electrochemical Society 163~(8) (2016) A1566--A1575.
\newblock \href {https://doi.org/10.1149/2.0441608jes}
  {\path{doi:10.1149/2.0441608jes}}.

\bibitem{Ai2020}
W.~Ai, L.~Kraft, J.~Sturm, A.~Jossen, B.~Wu,
  \href{https://iopscience.iop.org/article/10.1149/2.0122001JES}{{Electrochemical
  thermal-mechanical modelling of stress inhomogeneity in lithium-ion pouch
  cells}}, Journal of the Electrochemical Society 167~(1) (2020) 013512.
\newblock \href {https://doi.org/10.1149/2.0122001JES}
  {\path{doi:10.1149/2.0122001JES}}.

\bibitem{christensen2010}
J.~Christensen, {Modeling Diffusion-Induced Stress in Li-Ion Cells with Porous
  Electrodes}, Journal of The Electrochemical Society 157~(3) (2010) A366.
\newblock \href {https://doi.org/10.1149/1.3269995}
  {\path{doi:10.1149/1.3269995}}.

\bibitem{ebner2013}
M.~Ebner, F.~Marone, M.~Stampanoni, V.~Wood,
  \href{https://www.sciencemag.org/lookup/doi/10.1126/science.1241882}{{Visualization
  and Quantification of Electrochemical and Mechanical Degradation in Li Ion
  Batteries}}, Science 342~(6159) (2013) 716--720.
\newblock \href {https://doi.org/10.1126/science.1241882}
  {\path{doi:10.1126/science.1241882}}.

\bibitem{miehe2016}
C.~Miehe, H.~Dal, L.~M. Sch{\"{a}}nzel, A.~Raina,
  \href{http://doi.wiley.com/10.1002/nme.5133}{{A phase-field model for
  chemo-mechanical induced fracture in lithium-ion battery electrode
  particles}}, International Journal for Numerical Methods in Engineering
  106~(9) (2016) 683--711.
\newblock \href {https://doi.org/10.1002/nme.5133}
  {\path{doi:10.1002/nme.5133}}.

\bibitem{Klinsmann2016}
M.~Klinsmann, D.~Rosato, M.~Kamlah, R.~M. McMeeking,
  \href{http://dx.doi.org/10.1016/j.jmps.2016.04.004}{{Modeling crack growth
  during Li insertion in storage particles using a fracture phase field
  approach}}, Journal of the Mechanics and Physics of Solids 92 (2016)
  313--344.
\newblock \href {https://doi.org/10.1016/j.jmps.2016.04.004}
  {\path{doi:10.1016/j.jmps.2016.04.004}}.

\bibitem{Klinsmann2016a}
M.~Klinsmann, D.~Rosato, M.~Kamlah, R.~M. McMeeking, {Modeling Crack Growth
  during Li Extraction in Storage Particles Using a Fracture Phase Field
  Approach}, Journal of The Electrochemical Society 163~(2) (2016) A102--A118.
\newblock \href {https://doi.org/10.1149/2.0281602jes}
  {\path{doi:10.1149/2.0281602jes}}.

\bibitem{Mesgarnejad2019b}
A.~Mesgarnejad, A.~Karma,
  \href{https://linkinghub.elsevier.com/retrieve/pii/S0022509619305514}{{Phase
  field modeling of chemomechanical fracture of intercalation electrodes: Role
  of charging rate and dimensionality}}, Journal of the Mechanics and Physics
  of Solids 132 (2019) 103696.
\newblock \href {https://doi.org/10.1016/j.jmps.2019.103696}
  {\path{doi:10.1016/j.jmps.2019.103696}}.

\bibitem{xu2018}
R.~Xu, K.~Zhao, \href{https://doi.org/10.1016/j.jmps.2018.07.021}{{Corrosive
  fracture of electrodes in Li-ion batteries}}, Journal of the Mechanics and
  Physics of Solids 121 (2018) 258--280.
\newblock \href {https://doi.org/10.1016/j.jmps.2018.07.021}
  {\path{doi:10.1016/j.jmps.2018.07.021}}.

\bibitem{Paris1963}
P.~Paris, F.~Erdogan,
  \href{https://asmedigitalcollection.asme.org/fluidsengineering/article/85/4/528/395834/A-Critical-Analysis-of-Crack-Propagation-Laws}{{A
  Critical Analysis of Crack Propagation Laws}}, Journal of Basic Engineering
  85~(4) (1963) 528--533.
\newblock \href {https://doi.org/10.1115/1.3656900}
  {\path{doi:10.1115/1.3656900}}.

\bibitem{Purewal2014}
J.~Purewal, J.~Wang, J.~Graetz, S.~Soukiazian, H.~Tataria, M.~W. Verbrugge,
  \href{http://dx.doi.org/10.1016/j.jpowsour.2014.07.028}{{Degradation of
  lithium ion batteries employing graphite negatives and
  nickel-cobalt-manganese oxide + spinel manganese oxide positives: Part 2,
  chemical-mechanical degradation model}}, Journal of Power Sources 272 (2014)
  1154--1161.
\newblock \href {https://doi.org/10.1016/j.jpowsour.2014.07.028}
  {\path{doi:10.1016/j.jpowsour.2014.07.028}}.

\bibitem{ekstrom2015}
H.~Ekstr{\"{o}}m, G.~Lindbergh,
  \href{https://iopscience.iop.org/article/10.1149/2.0641506jes}{{A Model for
  Predicting Capacity Fade due to SEI Formation in a Commercial
  Graphite/LiFePO4 Cell}}, Journal of The Electrochemical Society 162~(6)
  (2015) A1003--A1007.
\newblock \href {https://doi.org/10.1149/2.0641506jes}
  {\path{doi:10.1149/2.0641506jes}}.

\bibitem{Laresgoiti2015}
I.~Laresgoiti, S.~K{\"{a}}bitz, M.~Ecker, D.~U. Sauer, {Modeling mechanical
  degradation in lithium ion batteries during cycling: Solid electrolyte
  interphase fracture}, Journal of Power Sources 300 (2015) 112--122.
\newblock \href {https://doi.org/10.1016/j.jpowsour.2015.09.033}
  {\path{doi:10.1016/j.jpowsour.2015.09.033}}.

\bibitem{Bourdin2000}
B.~Bourdin, G.~Francfort, J.~J. Marigo,
  \href{https://linkinghub.elsevier.com/retrieve/pii/S0022509699000289}{{Numerical
  experiments in revisited brittle fracture}}, Journal of the Mechanics and
  Physics of Solids 48~(4) (2000) 797--826.
\newblock \href {https://doi.org/10.1016/S0022-5096(99)00028-9}
  {\path{doi:10.1016/S0022-5096(99)00028-9}}.

\bibitem{Tanne2018}
E.~Tann{\'{e}}, T.~Li, B.~Bourdin, J.~J. Marigo, C.~Maurini,
  \href{https://linkinghub.elsevier.com/retrieve/pii/S0022509617306543}{{Crack
  nucleation in variational phase-field models of brittle fracture}}, Journal
  of the Mechanics and Physics of Solids 110 (2018) 80--99.
\newblock \href {https://doi.org/10.1016/j.jmps.2017.09.006}
  {\path{doi:10.1016/j.jmps.2017.09.006}}.

\bibitem{Quintanas-Corominas2020}
A.~Quintanas-Corominas, A.~Turon, J.~Reinoso, E.~Casoni, M.~Paggi, J.~Mayugo,
  \href{https://linkinghub.elsevier.com/retrieve/pii/S0045782519304943}{{A
  phase field approach enhanced with a cohesive zone model for modeling
  delamination induced by matrix cracking}}, Computer Methods in Applied
  Mechanics and Engineering 358 (2020) 112618.
\newblock \href {https://doi.org/10.1016/j.cma.2019.112618}
  {\path{doi:10.1016/j.cma.2019.112618}}.

\bibitem{Tan2021}
W.~Tan, E.~Mart{\'{i}}nez-Pa{\~{n}}eda,
  \href{https://linkinghub.elsevier.com/retrieve/pii/S0266353820323290}{{Phase
  field predictions of microscopic fracture and R-curve behaviour of
  fibre-reinforced composites}}, Composites Science and Technology 202 (2021)
  108539.
\newblock \href {https://doi.org/10.1016/j.compscitech.2020.108539}
  {\path{doi:10.1016/j.compscitech.2020.108539}}.

\bibitem{Hirshikesh2019}
{Hirshikesh}, S.~Natarajan, R.~K. Annabattula, E.~Mart{\'{i}}nez-Pa{\~{n}}eda,
  \href{https://linkinghub.elsevier.com/retrieve/pii/S135983681930229X}{{Phase
  field modelling of crack propagation in functionally graded materials}},
  Composites Part B: Engineering 169 (2019) 239--248.
\newblock \href {https://doi.org/10.1016/j.compositesb.2019.04.003}
  {\path{doi:10.1016/j.compositesb.2019.04.003}}.

\bibitem{Kumar2021}
A.~V.~P. Kumar, A.~Dean, J.~Reinoso, P.~Lenarda, M.~Paggi,
  \href{https://linkinghub.elsevier.com/retrieve/pii/S0263823120311046}{{Phase
  field modeling of fracture in Functionally Graded Materials:
  {\$}{\textbackslash}gamma{\$} -convergence and mechanical insight on the
  effect of grading}}, Thin-Walled Structures 159 (2021) 107234.
\newblock \href {https://doi.org/10.1016/j.tws.2020.107234}
  {\path{doi:10.1016/j.tws.2020.107234}}.

\bibitem{Simoes2021}
M.~Simoes, E.~Mart{\'{i}}nez-Pa{\~{n}}eda,
  \href{https://linkinghub.elsevier.com/retrieve/pii/S0045782520306897}{{Phase
  field modelling of fracture and fatigue in Shape Memory Alloys}}, Computer
  Methods in Applied Mechanics and Engineering 373 (2021) 113504.
\newblock \href {https://doi.org/10.1016/j.cma.2020.113504}
  {\path{doi:10.1016/j.cma.2020.113504}}.

\bibitem{FFEMS2022}
M.~Simoes, C.~Braithwaite, A.~Makaya, E.~Mart{\'{i}}nez-Pa{\~{n}}eda,
  \href{https://doi.org/10.1111/ffe.13638}{{Modelling fatigue crack growth in
  Shape Memory Alloys}}, Fatigue {\&} Fracture of Engineering Materials {\&}
  Structures 45 (2022) 1243--1257.
\newblock \href {https://doi.org/10.1111/ffe.13638}
  {\path{doi:10.1111/ffe.13638}}.

\bibitem{Wilson2016}
Z.~A. Wilson, C.~M. Landis,
  \href{https://linkinghub.elsevier.com/retrieve/pii/S002250961630285X}{{Phase-field
  modeling of hydraulic fracture}}, Journal of the Mechanics and Physics of
  Solids 96 (2016) 264--290.
\newblock \href {https://doi.org/10.1016/j.jmps.2016.07.019}
  {\path{doi:10.1016/j.jmps.2016.07.019}}.

\bibitem{Schuler2020}
L.~Schuler, A.~G. Ilgen, P.~Newell,
  \href{https://linkinghub.elsevier.com/retrieve/pii/S0045782520300190}{{Chemo-mechanical
  phase-field modeling of dissolution-assisted fracture}}, Computer Methods in
  Applied Mechanics and Engineering 362 (2020) 112838.
\newblock \href {https://doi.org/10.1016/j.cma.2020.112838}
  {\path{doi:10.1016/j.cma.2020.112838}}.

\bibitem{Abdollahi2012}
A.~Abdollahi, I.~Arias,
  \href{https://linkinghub.elsevier.com/retrieve/pii/S0022509612001354}{{Phase-field
  modeling of crack propagation in piezoelectric and ferroelectric materials
  with different electromechanical crack conditions}}, Journal of the Mechanics
  and Physics of Solids 60~(12) (2012) 2100--2126.
\newblock \href {https://doi.org/10.1016/j.jmps.2012.06.014}
  {\path{doi:10.1016/j.jmps.2012.06.014}}.

\bibitem{Martinez-Paneda2018}
E.~Mart{\'{i}}nez-Pa{\~{n}}eda, A.~Golahmar, C.~F. Niordson,
  \href{https://linkinghub.elsevier.com/retrieve/pii/S0045782518303529}{{A
  phase field formulation for hydrogen assisted cracking}}, Computer Methods in
  Applied Mechanics and Engineering 342 (2018) 742--761.
\newblock \href {https://doi.org/10.1016/j.cma.2018.07.021}
  {\path{doi:10.1016/j.cma.2018.07.021}}.

\bibitem{Anand2019}
L.~Anand, Y.~Mao, B.~Talamini,
  \href{https://linkinghub.elsevier.com/retrieve/pii/S0022509618303582}{{On
  modeling fracture of ferritic steels due to hydrogen embrittlement}}, Journal
  of the Mechanics and Physics of Solids 122 (2019) 280--314.
\newblock \href {https://doi.org/10.1016/j.jmps.2018.09.012}
  {\path{doi:10.1016/j.jmps.2018.09.012}}.

\bibitem{Wu2020a}
J.~Wu, T.~K. Mandal, V.~P. Nguyen,
  \href{https://linkinghub.elsevier.com/retrieve/pii/S0045782519304906}{{A
  phase-field regularized cohesive zone model for hydrogen assisted cracking}},
  Computer Methods in Applied Mechanics and Engineering 358 (2020) 112614.
\newblock \href {https://doi.org/10.1016/j.cma.2019.112614}
  {\path{doi:10.1016/j.cma.2019.112614}}.

\bibitem{Kristensen2020}
P.~K. Kristensen, E.~Mart{\'{i}}nez-Pa{\~{n}}eda,
  \href{https://doi.org/10.1016/j.tafmec.2019.102446
  https://linkinghub.elsevier.com/retrieve/pii/S0167844219305580}{{Phase field
  fracture modelling using quasi-Newton methods and a new adaptive step
  scheme}}, Theoretical and Applied Fracture Mechanics 107 (2020) 102446.
\newblock \href {https://doi.org/10.1016/j.tafmec.2019.102446}
  {\path{doi:10.1016/j.tafmec.2019.102446}}.

\bibitem{Mai2016}
W.~Mai, S.~Soghrati, R.~G. Buchheit,
  \href{https://linkinghub.elsevier.com/retrieve/pii/S0010938X16301408}{{A
  phase field model for simulating the pitting corrosion}}, Corrosion Science
  110 (2016) 157--166.
\newblock \href {https://doi.org/10.1016/j.corsci.2016.04.001}
  {\path{doi:10.1016/j.corsci.2016.04.001}}.

\bibitem{Cui2021}
C.~Cui, R.~Ma, E.~Mart{\'{i}}nez-Pa{\~{n}}eda,
  \href{https://linkinghub.elsevier.com/retrieve/pii/S0022509620304622}{{A
  phase field formulation for dissolution-driven stress corrosion cracking}},
  Journal of the Mechanics and Physics of Solids 147 (2021) 104254.
\newblock \href {https://doi.org/10.1016/j.jmps.2020.104254}
  {\path{doi:10.1016/j.jmps.2020.104254}}.

\bibitem{Li2021}
W.~Li, K.~Shirvan,
  \href{https://linkinghub.elsevier.com/retrieve/pii/S0272884220325712}{{Multiphysics
  phase-field modeling of quasi-static cracking in urania ceramic nuclear
  fuel}}, Ceramics International 47~(1) (2021) 793--810.
\newblock \href {https://doi.org/10.1016/j.ceramint.2020.08.191}
  {\path{doi:10.1016/j.ceramint.2020.08.191}}.

\bibitem{Bourdin2008}
B.~Bourdin, G.~A. Francfort, J.-J. Marigo,
  \href{http://link.springer.com/10.1007/s10659-007-9107-3}{{The Variational
  Approach to Fracture}}, Journal of Elasticity 91~(1-3) (2008) 5--148.
\newblock \href {https://doi.org/10.1007/s10659-007-9107-3}
  {\path{doi:10.1007/s10659-007-9107-3}}.

\bibitem{Kristensen2021}
P.~K. Kristensen, C.~F. Niordson, E.~Mart{\'{i}}nez-Pa{\~{n}}eda,
  \href{https://royalsocietypublishing.org/doi/10.1098/rsta.2021.0021}{{An
  assessment of phase field fracture: crack initiation and growth}},
  Philosophical Transactions of the Royal Society A: Mathematical, Physical and
  Engineering Sciences 379~(2203) (2021) 20210021.
\newblock \href {https://doi.org/10.1098/rsta.2021.0021}
  {\path{doi:10.1098/rsta.2021.0021}}.

\bibitem{Miehe2010}
C.~Miehe, F.~Welschinger, M.~Hofacker,
  \href{http://doi.wiley.com/10.1002/nme.2861}{{Thermodynamically consistent
  phase-field models of fracture: Variational principles and multi-field FE
  implementations}}, International Journal for Numerical Methods in Engineering
  83~(10) (2010) 1273--1311.
\newblock \href {https://doi.org/10.1002/nme.2861}
  {\path{doi:10.1002/nme.2861}}.

\bibitem{Ambati2014}
M.~Ambati, T.~Gerasimov, L.~De~Lorenzis, {A review on phase-field models of
  brittle fracture and a new fast hybrid formulation}, Computational Mechanics
  55~(2) (2014) 383--405.
\newblock \href {https://doi.org/10.1007/s00466-014-1109-y}
  {\path{doi:10.1007/s00466-014-1109-y}}.

\bibitem{Amor2009}
H.~Amor, J.-J. Marigo, C.~Maurini,
  \href{https://linkinghub.elsevier.com/retrieve/pii/S0022509609000659}{{Regularized
  formulation of the variational brittle fracture with unilateral contact:
  Numerical experiments}}, Journal of the Mechanics and Physics of Solids
  57~(8) (2009) 1209--1229.
\newblock \href {https://doi.org/10.1016/j.jmps.2009.04.011}
  {\path{doi:10.1016/j.jmps.2009.04.011}}.

\bibitem{Lo2019}
Y.-S. Lo, M.~J. Borden, K.~Ravi-Chandar, C.~M. Landis,
  \href{https://linkinghub.elsevier.com/retrieve/pii/S0022509619306568}{{A
  phase-field model for fatigue crack growth}}, Journal of the Mechanics and
  Physics of Solids 132 (2019) 103684.
\newblock \href {https://doi.org/10.1016/j.jmps.2019.103684}
  {\path{doi:10.1016/j.jmps.2019.103684}}.

\bibitem{Carrara2020}
P.~Carrara, M.~Ambati, R.~Alessi, L.~De~Lorenzis,
  \href{https://doi.org/10.1016/j.cma.2019.112731
  https://linkinghub.elsevier.com/retrieve/pii/S0045782519306218}{{A framework
  to model the fatigue behavior of brittle materials based on a variational
  phase-field approach}}, Computer Methods in Applied Mechanics and Engineering
  361 (2020) 112731.
\newblock \href {https://doi.org/10.1016/j.cma.2019.112731}
  {\path{doi:10.1016/j.cma.2019.112731}}.

\bibitem{Schreiber2020}
C.~Schreiber, C.~Kuhn, R.~M{\"{u}}ller, T.~Zohdi,
  \href{https://doi.org/10.1007/s10704-020-00468-w}{{A phase field modeling
  approach of cyclic fatigue crack growth}}, International Journal of Fracture
  225~(1) (2020) 89--100.
\newblock \href {https://doi.org/10.1007/s10704-020-00468-w}
  {\path{doi:10.1007/s10704-020-00468-w}}.

\bibitem{Stallard2022MechanicalBatteries}
J.~C. Stallard, L.~Wheatcroft, S.~G. Booth, R.~Boston, S.~A. Corr, M.~F.
  De~Volder, B.~J. Inkson, N.~A. Fleck, {Mechanical properties of cathode
  materials for lithium-ion batteries}, Joule 6~(5) (2022) 984--1007.
\newblock \href {https://doi.org/10.1016/j.joule.2022.04.001}
  {\path{doi:10.1016/j.joule.2022.04.001}}.

\bibitem{boyce2022}
A.~M. Boyce, E.~Mart{\'{i}}nez-Pa{\~{n}}eda, A.~Wade, Y.~S. Zhang, J.~J.
  Bailey, T.~M. Heenan, D.~J. Brett, P.~R. Shearing, {Cracking predictions of
  lithium-ion battery electrodes by X-ray computed tomography and modelling},
  Journal of Power Sources 526 (2022) 231119.
\newblock \href {https://doi.org/10.1016/j.jpowsour.2022.231119}
  {\path{doi:10.1016/j.jpowsour.2022.231119}}.

\bibitem{Severson2019}
K.~A. Severson, P.~M. Attia, N.~Jin, N.~Perkins, B.~Jiang, Z.~Yang, M.~H. Chen,
  M.~Aykol, P.~K. Herring, D.~Fraggedakis, M.~Z. Bazant, S.~J. Harris, W.~C.
  Chueh, R.~D. Braatz, {Data-driven prediction of battery cycle life before
  capacity degradation}, Nature Energy 4~(5) (2019) 383--391.
\newblock \href {https://doi.org/10.1038/s41560-019-0356-8}
  {\path{doi:10.1038/s41560-019-0356-8}}.

\bibitem{J.R.Rice1968}
J.~R. Rice,
  \href{https://asmedigitalcollection.asme.org/appliedmechanics/article/35/2/379/392117/A-Path-Independent-Integral-and-the-Approximate}{{A
  Path Independent Integral and the Approximate Analysis of Strain
  Concentration by Notches and Cracks}}, Journal of Applied Mechanics 35~(2)
  (1968) 379--386.
\newblock \href {https://doi.org/10.1115/1.3601206}
  {\path{doi:10.1115/1.3601206}}.

\bibitem{chen2020a}
C.-H. Chen, F.~Brosa~Planella, K.~O’Regan, D.~Gastol, W.~D. Widanage,
  E.~Kendrick,
  \href{https://iopscience.iop.org/article/10.1149/1945-7111/ab9050}{{Development
  of Experimental Techniques for Parameterization of Multi-scale Lithium-ion
  Battery Models}}, Journal of The Electrochemical Society 167~(8) (2020)
  080534.
\newblock \href {https://doi.org/10.1149/1945-7111/ab9050}
  {\path{doi:10.1149/1945-7111/ab9050}}.

\bibitem{Liu2012}
X.~H. Liu, L.~Zhong, S.~Huang, S.~X. Mao, T.~Zhu, J.~Y. Huang, {Size-dependent
  fracture of silicon nanoparticles during lithiation}, ACS Nano 6~(2) (2012)
  1522--1531.
\newblock \href {https://doi.org/10.1021/nn204476h}
  {\path{doi:10.1021/nn204476h}}.

\bibitem{Ebner2013b}
M.~Ebner, F.~Geldmacher, F.~Marone, M.~Stampanoni, V.~Wood, {X-Ray Tomography
  of Porous, Transition Metal Oxide Based Lithium Ion Battery Electrodes},
  Advanced Energy Materials 3~(7) (2013) 845--850.
\newblock \href {https://doi.org/10.1002/aenm.201200932}
  {\path{doi:10.1002/aenm.201200932}}.

\bibitem{Lu2020NC}
X.~Lu, A.~Bertei, D.~P. Finegan, C.~Tan, S.~R. Daemi, J.~S. Weaving, K.~B.
  O’Regan, T.~M.~M. Heenan, G.~Hinds, E.~Kendrick, D.~J.~L. Brett, P.~R.
  Shearing, \href{http://dx.doi.org/10.1038/s41467-020-15811-x
  http://www.nature.com/articles/s41467-020-15811-x}{{3D microstructure design
  of lithium-ion battery electrodes assisted by X-ray nano-computed tomography
  and modelling}}, Nature Communications 11~(1) (2020) 2079.
\newblock \href {https://doi.org/10.1038/s41467-020-15811-x}
  {\path{doi:10.1038/s41467-020-15811-x}}.

\bibitem{Zhao2016b}
Y.~Zhao, B.~Xu, P.~Stein, D.~Gross,
  \href{https://linkinghub.elsevier.com/retrieve/pii/S0045782516302845}{{Phase-field
  study of electrochemical reactions at exterior and interior interfaces in
  Li-ion battery electrode particles}}, Computer Methods in Applied Mechanics
  and Engineering 312 (2016) 428--446.
\newblock \href {https://doi.org/10.1016/j.cma.2016.04.033}
  {\path{doi:10.1016/j.cma.2016.04.033}}.

\bibitem{DFN1993}
M.~Doyle, T.~F. Fuller, J.~Newman,
  \href{https://iopscience.iop.org/article/10.1149/1.2221597}{{Modeling of
  Galvanostatic Charge and Discharge of the Lithium/Polymer/Insertion Cell}},
  Journal of The Electrochemical Society 140~(6) (1993) 1526--1533.
\newblock \href {https://doi.org/10.1149/1.2221597}
  {\path{doi:10.1149/1.2221597}}.

\bibitem{Gao2013a}
Y.~Gao, M.~Zhou,
  \href{https://linkinghub.elsevier.com/retrieve/pii/S0378775312018800}{{Coupled
  mechano-diffusional driving forces for fracture in electrode materials}},
  Journal of Power Sources 230 (2013) 176--193.
\newblock \href {https://doi.org/10.1016/j.jpowsour.2012.12.034}
  {\path{doi:10.1016/j.jpowsour.2012.12.034}}.

\end{thebibliography}

\end{document}